\documentclass[fleqn,usenatbib]{mnras}

\usepackage{newtxtext,newtxmath}

\usepackage[T1]{fontenc}
\usepackage{ae,aecompl}

\usepackage{graphicx}	
\usepackage{amsmath}	
\usepackage{amssymb}	

\usepackage[dvipsnames]{xcolor} 
\definecolor{orange}{rgb}{1,0.5,0}
\usepackage[normalem]{ulem} 

\usepackage{xspace} 
\newcommand{\sage}{\texttt{SAGE}\xspace}
\newcommand{\Msun}{\ensuremath{\mathrm{M}_{\odot}}\xspace}
\newcommand{\Zsun}{\ensuremath{\mathrm{Z}_{\odot}}\xspace}

\usepackage{accents} 
\newcommand*{\dt}[1]{%
  \accentset{\mbox{\large\bfseries .}}{#1}}

\title[The origin of dust in galaxies across cosmic time]{The origin of dust in galaxies across cosmic time}

\author[D. P. Triani et al.]{
Dian P. Triani,$^{1,2}$\thanks{E-mail: dtriani@swin.edu.au}
Manodeep Sinha,$^{1,2}$
Darren J. Croton,$^{1,2}$
Camilla Pacifici,$^{3}$
\newauthor
and Eli Dwek$^{4}$
\\
$^{1}$Centre for Astrophysics \& Supercomputing, Swinburne University of Technology, Hawthorn, VIC 3122, Australia\\
$^{2}$ARC Centre of Excellence for All Sky Astrophysics in 3 Dimensions (ASTRO 3D)\\
$^{3}$Space Telescope Science Institute, Baltimore, MD 21218, USA\\
$^{4}$Observational Cosmology Lab, NASA Goddard Space Flight Center, Code 665, Greenbelt, MD 20771, USA\\
}

\date{Accepted XXX. Received YYY; in original form ZZZ}

\pubyear{2019}

\hypersetup{draft}
\begin{document}
\label{firstpage}
\pagerange{\pageref{firstpage}--\pageref{lastpage}}
\maketitle

\begin{abstract}
We study the dust evolution in galaxies by implementing a detailed dust prescription in the \texttt{SAGE} semi-analytical model for galaxy formation. The new model, called \texttt{Dusty SAGE}, follows the condensation of dust in the ejecta of type II supernovae and asymptotic giant branch (AGB) stars, grain growth in the dense molecular clouds, destruction by supernovae shocks, and the removal of dust from the ISM by star formation, reheating, inflows and outflows. Our model successfully reproduces the observed dust mass function at redshift $z=0$ and the observed scaling relations for dust across a wide range of redshifts. We find that the dust mass content in the present Universe is mainly produced via grain growth in the interstellar medium (ISM). By contrast, in the early Universe, the primary production mechanism for dust is the condensation in stellar ejecta. The shift of the significant production channel for dust characterises the scaling relations of dust-to-gas (DTG) and dust-to-metal (DTM) ratios. In galaxies where the grain growth dominates, we find positive correlations for DTG and DTM ratios with both metallicity and stellar mass. On the other hand, in galaxies where dust is produced primarily via condensation, we find negative or no correlation for DTM and DTG ratios with either metallicity or stellar mass. In agreement with observation showing that the circumgalactic medium (CGM) contains more dust than the ISM, our model also shows the same trend for $z < 4$. Our semi-analytic model is publicly available at \url{https://github.com/dptriani/dusty-sage}.

\end{abstract}



\section{Introduction}

Galaxies are complicated systems that comprise many components, including stars, gas, dust and metals. Although stars are the primary source of galactic optical emission, the other components significantly affect the spectral energy distribution (SED) in the other wavelength regimes. Dust is one of the key components that affects the galactic SED over a broad range of wavelengths. Dust absorbs most of the ultraviolet light from stars and reemits the radiation in the infrared \citep{WTC92, WG00}. The internal properties of dust shape the distribution of galaxy light across much of the wavelength range \citep{Granato00}.

Aside from altering the observed SED, dust also plays many significant roles in the physics of galaxy evolution, especially at early stages \citep[e.g.,][]{Yamasawa11}. Dust acts as an efficient cooling channel for gas in the hot halo \citep{OS73, Dwek87}, a depletion agent for metals and a catalyst for the formation of molecular hydrogen in the interstellar medium (ISM) \citep{HMK79}. As molecular hydrogen is the raw material for stars, star formation activity in the ISM is enabled by the existence of dust.

One way to assess the impact of dust on galaxy evolution is by implementing a detailed dust model within a cosmological simulation. Such cosmological simulations follow the formation of large scale structure and galaxies that we observe today, starting from a nearly uniform density field at early times. The two common approaches for simulating galaxy formation are semi-analytic models (SAMs) and hydrodynamic simulations. Semi-analytic models evolve baryons in dark matter halos provided by either the Press-Schechter formalism or an N-body simulation \citep{WF91, Croton06, Somerville08, Croton16}. Hydrodynamic simulations evolve the dark matter and baryons simultaneously, resulting in a precise but computationally expensive calculation (\citealp[e.g., Illustris,][]{Vogelsberger14}; \citealp[EAGLE,][]{Schaye15}).

Before implementing a dust model in a galaxy evolution simulation, we first need to carefully track the metal abundance as metal grains are the raw material for dust formation
in galaxies \citep[e.g.,][]{DS80}. \citet{Arrigoni10} include an enhanced metal enrichment model in the updated \texttt{Santa Cruz} semi-analytic model \citep{Somerville08} to predict the abundance of individual elements in galaxies. Their model predictions match the observed mass-metallicity relation and the trend of increasing abundance ratio $[\alpha/\mathrm{Fe}]$ with galactic stellar mass.

Interstellar dust is produced by the condensation of metals in the stellar ejecta \citep[e.g.,][]{Dwek1998, ZGT08}. Low mass stars eject most of their metals during an asymptotic giant branch phase \citep[AGB,][]{Valiante09} while higher mass stars undergo either a type II supernovae (SN II) or a type Ia supernovae (SN Ia) phase to expel most of their metal content. However, there is currently no observational evidence for the formation of dust in the ejecta of SN Ia \citep{Dwek16}. After the initial formation, dust grains grow via accretion in the ISM \citep[e.g.,][]{Dwek1998, ZGT08, Draine09}. There is still debate as to which production channel produces more dust in the ISM. SN shocks are the main destroyers of dust in the ISM \citep{ZGT08, Slavin15}, with grain-grain collisions and thermal sputtering as the main destruction mechanisms in the hot halo and ejected reservoir.

Despite the importance of dust in galaxies, the majority of cosmological simulations and models only include a simplified accounting of dust. The standard approach is to model ISM enrichment by assuming a constant yield for metals in every star formation episode \citep{Baugh05, Somerville12, Croton16}. More detailed studies of dust in galaxy evolution are made using so-called ``closed-box" models \citep[e.g.,][]{Dwek1998, ZGT08, Valiante09, Asano13}. These models are specifically made to investigate the formation and destruction processes of dust. However, they do not typically focus on the properties of galaxies within a cosmological context.

The first effort to study dust in a hydrodynamical simulation was \citet{McKinnon16}. They coupled a detailed dust model to the moving-mesh code \texttt{AREPO}. The model included stellar production of dust, grain growth, and destruction by supernovae to investigate dust formation in eight Milky-Way sized haloes. \citet{McKinnon17} extended the model to include a prescription for thermal sputtering and simulated a more diverse galaxy population. In \citet{McKinnon18}, the model was further extended to account for the evolution of the grain size distribution. 

\citet{Popping17} was one of the first semi-analytic models that included a detailed dust model. They extended the work by \citet{Arrigoni10} and added prescriptions for dust formation and destruction in the ISM and hot halo. Recently, \citet{Vijayan19} implemented a detailed metal and dust enrichment model to the latest version of the popular \texttt{L-GALAXIES} semi analytic model \citep[][and references therein]{Henriques15}.

Building on previous works, this paper describes a new model for dust and metal enrichment, coupled to the Semi Analytic Galaxy Evolution \citep[\texttt{SAGE}\footnote{https://github.com/darrencroton/sage},][]{Croton16} model. \texttt{SAGE} builds galaxies using a set of empirical and analytical prescriptions, taking a full set of dark halo merger trees from an N-body cosmological simulation as input. Our version is called \texttt{Dusty SAGE}\footnote{https://github.com/dptriani/dusty-sage} and retains the modular design of \sage.

Our model differs from the work of \citet{Vijayan19} in the treatment of dust beyond the ISM. Their model assumes that dust blown out of the galaxy disc is always destroyed. However, considering the massive amount of dust observed in the circumgalactic medium (CGM), we follow the approach of \citet{Popping17} that include dust outflows to the hot halo and ejected reservoir. A fraction of dust that survives the thermal sputtering in the hot halo and ejected reservoir can be re-accreted back to the ISM. The main difference between our model and \citet{Popping17} is that our model runs on a set of N-body simulation halo merger trees such as those from the Millennium simulation \citep{Springel05}, while theirs used the trees from an extended Press-Schechter formalism described in \citet{1999SK}. Our model also differs from previous models in the treatment of dust SN Ia ejecta. Most previous models formed dust grains from the condensation of metals in the SN Ia ejecta, while recent studies have showed the lack of such dust \citep{Gomez12, Dwek16}. This absence is likely due to the low mass and high velocities materials ejected in the SN Ia compared to those of SN II, and the short time allowed by SN Ia for dust grains to be formed because of the drop of temperature and density after the SN Ia event. Considering this, our model produces heavy elements in SN Ia but does not allow them to condense into dust.

This paper is organised as follows. In Section \ref{sec:galevo}, we describe the \sage semi-analytic model and the enhancements we have implemented, including the differentiation of cold gas into molecular and atomic gas and a detailed metal abundance tracking. In Section \ref{sec:dustevo} we introduce our new dust prescription. The results and discussion are presented in Section \ref{sec:result}, and conclusions given in Section \ref{sec:conclusion}. Throughout this paper, we adopt the \citet{Chabrier03} initial mass function and assume $h = 0.73$ based on the cosmology used for the Millennium simulation.

\section{Galaxy evolution model}

\begin{figure*}
    \centering
    \includegraphics[width=0.8\textwidth]{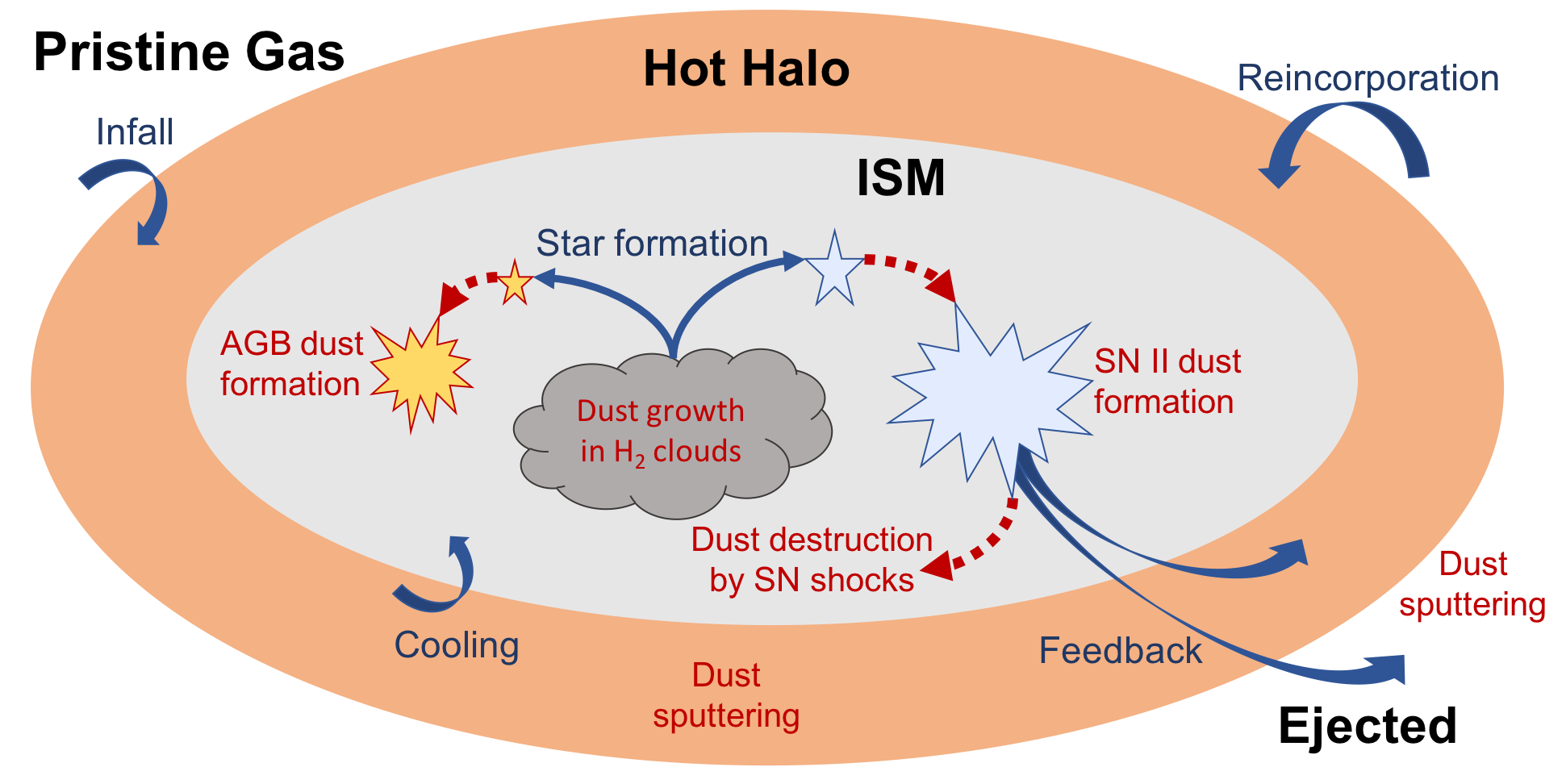}
    \caption{A schematic diagram of baryonic and dust processes in our galaxy evolution model, \texttt{Dusty SAGE}. The baryon reservoirs are marked with thick black fonts. The galaxy evolution processes between these reservoirs is marked with blue arrows and blue fonts. The dust physics in marked with red dashed arrows and red fonts. Details are provided in text (Section \ref{sec:galevo})}.
    \label{fig:scheme}
\end{figure*}
\label{sec:galevo}

In this section, we describe the galaxy evolution model, \texttt{SAGE}, that we use as a base for our new dust model. We present an overview of \texttt{SAGE} (Section~\ref{sec:sage}) then discuss the changes made for the present work in \texttt{Dusty SAGE}. These changes include a new star formation prescription, the calculation of the molecular gas fraction (Section~\ref{sec:mc_ism}), and a self-consistent chemical enrichment model (Section~\ref{sec:chem_enrich}).

Figure \ref{fig:scheme} shows the backbone of \texttt{Dusty SAGE}. We trace movement of baryons (gas, stellar, metals) in pristine gas, hot halo, ISM, and ejected reservoirs. The flow of dust throughout these reservoirs is overlayed showing the various dust creation and destruction process that comprises our detailed dust model.

\subsection{The Semi Analytic Model - SAGE}\label{sec:sage}
\texttt{SAGE} follows the evolution of baryons by using dark matter halo merger trees as an input and solving a set of simplified differential equations. Each equation represents a critical process in galaxy evolution, including: infall of the pristine gas to the hot halo; gas cooling into the disk; star formation in the ISM; SN and AGN feedback processes that reheat and eject baryons from the ISM back to the hot and ejected reservoirs; and the reincorporation of ejected baryons back to the hot halo (see Figure \ref{fig:scheme}). The halo merger trees are taken from the Millennium simulation \citep{Springel05}. This method is inexpensive in terms of computing time relative to hydrodynamical simulations.
The predecessor of \texttt{SAGE} was first presented in \citet{Croton06}. This model was tuned to reproduce the observed stellar mass function at redshift z=0 by introducing two modes in a new active galactic nuclei (AGN) feedback prescription: the radio mode and the quasar mode. \citet{Croton16} updated the galaxy model by adding more physics, and publicly released it as \texttt{SAGE}. The code is written in \texttt{C} and is optimised to be efficient, modular and adaptive. \sage can produce sensible results on a broad range of dark matter simulations with only minor adjustments to the free parameters. Several authors have adapted the code and implemented different physics suited to their research needs. \citet{Stevens16} added radial tracking for both atomic and molecular hydrogen in the galaxy disc in a version called \texttt{DARK SAGE}\footnote{https://github.com/arhstevens/DarkSage}. \citet{2017Raouf} expanded the AGN jet physics available in a version called \texttt{Radio SAGE}. While \citet{Seiler19} coupled \texttt{SAGE} with a hydrogen ionisation code to investigate the epoch of reionisation in a version named \texttt{RSAGE} \footnote{https://github.com/jacobseiler/rsage}.

In this work, we use the 2016 version of \texttt{SAGE} \citep{Croton16} as a base and add the following enhancements: the differentiation of cold gas in the ISM into HI and $\mathrm{H_2}$; a detailed chemical enrichment model using AGB stars, SNIa and SNII; and a new self-consistent dust evolution model. We also update the star formation prescription to use molecular hydrogen as the raw material for stars. Unless otherwise stated, the remaining model and parameters are unchanged from those described in \citet{Croton16}.

\subsection{Molecular clouds in the ISM}\label{sec:mc_ism}
Cold gas in the ISM exists in two phases -- giant molecular clouds and diffuse gas. Such giant molecular clouds are the primary sites for star formation activity. We follow the procedure described in \citet{Stevens16} to determine the fraction of molecular-to-atomic gas in the ISM and to invoke star formation from molecular clouds. However, we apply the procedure to the entire galactic disc, not to the defined annuli in the disc that was the focus of \citet{Stevens16}. 

An empirical study by \citet{Leroy08} found a tight relation between the surface density of star formation and the surface density of molecular clouds \citep{Kennicutt12}. This relation has become the basis for the star formation prescription in many semi-analytic models \citep[e.g.,][]{Stevens16, Fu13}:

\begin{equation}
    \Sigma_\mathrm{SFR} = \epsilon_\mathrm{SF} \Sigma_\mathrm{H_2} \; ,
    \label{eq:1}
\end{equation}

where $\epsilon_\mathrm{SF}$ is the star formation efficiency and $\Sigma_\mathrm{H_2}$ is the surface density of molecular hydrogen. In our model $\epsilon_\mathrm{SF}$ is taken as a bound free parameter with a default value $\epsilon_\mathrm{SF} = 0.005 \mathrm{Myr}^{-1}$.

To compute the surface density of the molecular hydrogen, we first break the cold gas into atomic and molecular components. We use the formula proposed by \citet{BR04, BR06} to determine the ratio of molecular to atomic hydrogen in the galactic disk using mid-plane pressures, expressed as:

\begin{equation}
    R_\mathrm{H_2} = \frac{\Sigma_\mathrm{H_2}}{\Sigma_\mathrm{HI}} = \left[\frac{P}{P_0}\right]^{\chi},
    \label{eq:ratio}
\end{equation}
where $P_0 = 5.93\times10^{-13} h^2$ Pa ($P_0/k = 43000\ \mathrm{cm}^{-3}\ \mathrm{K}$) and $\chi = 0.92$ \citep{BR06}. According to \citet{Elmegreen89, Elmegreen93}, the mid-plane pressure can be written:
\begin{equation}
    P = \frac{\pi}{2} G \Sigma_\mathrm{gas} [\Sigma_\mathrm{gas} + f_\mathrm{\sigma} \Sigma_*] \; ,
    \label{eq:2}
\end{equation}
where $G$ is the gravitational constant, $\Sigma_\mathrm{gas}$ is the surface density of cold gas, $\Sigma_\mathrm{*}$ is the surface density of stars and $f_\mathrm{\sigma}$ is the ratio of the velocity dispersions of the gas and stars. We adopt an average value of $f_\mathrm{\sigma}$ from \citet{Elmegreen93}, $f_\mathrm{\sigma} = 0.4$. 
We use one representative gas surface density value to calculate $R_\mathrm{H_2}$ for each galaxy. Following the Milky Way as an example, we assume that the size of the disk is 3 times its scale radius and the surface density is calculated using this disk size. The scale radius is derived from the viral radius and spin parameter of the dark halo following \citet{1998MMW}.

Diffuse gas in the ISM contains atomic hydrogen, metals and ionised hydrogen. As the formula above only computes the fraction of the molecular to atomic hydrogen instead of the fraction of the molecular hydrogen to the total cold gas mass, we need a correction. In particular, the cold gas mass not only consists of molecular and atomic hydrogen, but also contains warm ionized hydrogen, helium, metals and dust. As stated in \citet{Stevens16}, the true fraction of $\mathrm{H_2}$ can be expressed as:
\begin{equation}
    f_\mathrm{H_2} = \frac{\Sigma_\mathrm{H_2}}{\Sigma_\mathrm{gas}} = \frac{f_\mathrm{He} \ f_\mathrm{warm}}{R_\mathrm{H_2}^{-1} + 1} [1 - Z_\mathrm{gas}] \; ,
    \label{eq:trueration}
\end{equation}
where $Z_\mathrm{gas}$ is the metallicity of the gas. We adopt $f_\mathrm{He} = 0.75$ and $f_\mathrm{warm} = 1/1.3$ from \citet{Fu10}.

\subsection{Chemical enrichment model}\label{sec:chem_enrich}
Like many semi analytic models, \texttt{SAGE} uses an instantaneous recycling approximation to model the chemical enrichment in galaxies. In such SAMs, metals are considered to form only in an SN II event and are assumed to have a constant yield in every star formation episode. By contrast, in this work, we consider the enrichment from AGB, SN II and SN Ia and make use of the stellar yields.

We follow \citet{Arrigoni10} to compute the total rate of element $\dt{M_j}$ ejected into the ISM at a time $t$:

\begin{equation} \label{eq:5}
\begin{split}
       \dt{M_j} &= \int_{M_L}^{M_{B_m}} \psi(t-\tau_M) Q_{j}^\mathrm{AGB}(t-\tau_M) \phi(M) dM \\ 
       & + (1-A) \int_{M_{B_m}}^{M_{AGB}} \psi(t-\tau_M) Q_{j}^\mathrm{AGB}(t-\tau_M) \phi(M) dM \\
       & + A \int_{M_{B_m}}^{M_{B_M}}\phi(M_B)  \\
       & \times \left[\int_{\mu_\mathrm{min}}^{0.5} f(\mu) \psi(t-\tau_{M_2}) Q_{j}^\mathrm{SNIa}(t-\tau_{M_2}) d\mu \right] dM_B     \\
       & + (1-A) \int_{M_{SNII}}^{M_{B_M}} \psi(t-\tau_M) Q_{j}^\mathrm{SNII}(t-\tau_M) \phi(M) dM \\
       & + \int_{M_{B_M}}^{M_U} \psi(t-\tau_M) Q_{j}^\mathrm{SNII}(t-\tau_M) \phi(M) dM.
\end{split}
\end{equation}
Here each term represents a different stellar source that contributes to the total metal abundance in the ISM. $\phi(M)$ is the initial mass function (IMF). $\psi(t-\tau_M)$ is the star formation rate at an epoch $(t-\tau_M)$, where $\tau_M$ is the main sequence lifetime of a star of mass $M$. We compute $\tau_M$ using Equation 3 from \citet{RVR96}. $\mu$ is mass fraction of the secondary star in binary systems $(\mu = M_2/M_B)$ and $f(\mu)$ is the distribution of this fraction. Following \citet{Arrigoni10}, the distribution function of the secondary mass fraction in the binary pair can be defined as
\begin{equation}\label{eq:7}
    f(\mu) = 2^{1+\gamma}(1+\gamma)\mu^\gamma,
\end{equation}
with $\gamma=2$. $Q_{j}^\mathrm{AGB}(t)$, $Q_{j}^\mathrm{SNIa}(t)$, and $Q_{j}^\mathrm{SNII}(t)$ are the yields of elements ejected into ISM by the AGB winds and the explosion of SNIa and SNII, respectively.

The first and second integrals in Equation \ref{eq:5} represent the contribution of single stars with low stellar mass $(1-8 M_\odot)$ that produce metals via AGB winds. $M_L = 0.1 M_\odot$ is the assumed lower mass limit of stars formed, $M_{B_m} = 3 \Msun$ is the minimum mass of a binary pair, and $M_{AGB} = 8 \Msun$ is the upper mass of stars that will end their lives on the AGB. Between $M_{B_m}$ and $M_{B_M} = 16 \Msun$, a fraction of stars consist of binary pairs. We assume all binary stars end up as SN Ia. The parameter $A$ represents this fraction and we use $A=0.04$ under the constraint set by the chemical evolution model for the Milky Way of \citet{Francois04}. The third integral represents the contribution of binary pairs that eject metals into the ISM via SNIa explosions. In this case, the timing of the explosion is set up by the lifetime of the secondary star. The final two integrals represent the contribution of single stars with mass $M_{SNII} = 8 \Msun$ and $M_{U} = 40 \Msun$. These stars contribute metals via the ejecta of SN II. 

We use the IMF of \citet{Chabrier03}:
\begin{equation} \label{eq:6}
  f(n) = \begin{cases} C_1 e^{-(\log m - \log m_c)^2 / 2 \sigma^2} &\mbox{if } m < 1 M_\odot. \\ 
  (C_2 m^{-x} & \mbox{if } m > 1 M_\odot. \end{cases}
\end{equation}
We adopt the standard parameters of the IMF, $m_c = 0.079 M_\odot$, $\sigma = 0.69$ and $x = 1.3$. We normalise this function in the mass interval $0.1 - 40 M_\odot$ to contrain the constant values of $C_1 = 0.9098$ and $C_2 = 0.2539$.

Stellar yield ($Q_{j}$ in the equation above) is the amount of an element ejected by stars at the end of their lives. We adopt different nucleosynthesis grids for different stellar mass ranges. For AGB stars with mass range $1 < M / M_\odot < 8$ and metallicities $Z=(0.0, 1\times10^{-4}, 4\times10^{-4}, 0.004, 0.008, 0.02, 0.05)$, we track C, N and O using \citet{Karakas10}. For SN II stars with mass $M > 8 M_\odot$ and the same metallicity grid as above, we use the yield values for C, O, Mg, Si, S, Ca and Fe taken from \citet{WW95}. For SN Ia stars, we adopt yields for Cr, Fe and Ni from \citet{Seitenzahl13}, assuming solar metallicity. 

\section{Dust Evolution Model}
\label{sec:dustevo}
Dust evolves in galaxies via many complex physical processes. In this section, we explain the processes that we include in our model and how we implement them: the initial formation in stellar ejecta (Section 3.1), growth in molecular clouds (Section 3.2), destruction by SNe (Section 3.3), other processes in the ISM (Section 3.4) and thermal sputtering of dust grains in the hot halo (Section 3.5). The summary of these processes are shown as red dashed arrows and red fonts in Figure \ref{fig:scheme}. The model is built in the same manner as \citet{Popping17}. 

The overall equation that describes the evolution of dust in our model galaxies is:

\begin{equation} \label{eq:8}
    \dt{M}_\mathrm{d} = \dt{M}_\mathrm{d}^\mathrm{form} + \dt{M}_\mathrm{d}^\mathrm{growth} - \dt{M}_\mathrm{d}^\mathrm{dest} - \dt{M}_\mathrm{d}^\mathrm{SF} - \dt{M}_\mathrm{d}^\mathrm{outflow} + \dt{M}_\mathrm{d}^\mathrm{inflow},
\end{equation}
where $\dt{M}_\mathrm{d}^\mathrm{form}$ is the rate of dust formation in stellar sources (AGB and SN II), $\dt{M}_\mathrm{d}^\mathrm{growth}$ is the rate of dust growth in molecular clouds, $\dt{M}_\mathrm{d}^\mathrm{dest}$ is the rate of dust destruction by SN, $\dt{M}_\mathrm{d}^\mathrm{SF}$ is the rate of dust that goes back to stars in every star formation episode, $\dt{M}_\mathrm{d}^\mathrm{outflow}$ is the rate of dust that is ejected out of the ISM by the feedback processes, and $\dt{M}_\mathrm{d}^\mathrm{infall}$ is the rate that dust is accreted back to the ISM. We exchange masses between the dust and metal reservoirs during these processes: when some dust mass is formed, the same amount of mass is removed from the metals reservoir, and when an amount of dust is destroyed, same amount of mass is added back into the metals reservoir. 

\subsection{Initial formation in stellar ejecta}

Stars that are born in pristine molecular clouds consist mainly of hydrogen. Nuclear reactions in the stellar core then turn hydrogen into heavier elements (``metals"). When stars die, these elements pollute the ISM and condense into dust. Similar to \citet{Popping17}, we follow the \citet{Dwek1998} approach to model the dust condensation in stellar ejecta. 

For AGB stars, we track the abundances of C, N and O in every star formation episode. The total dust mass that is formed by the condensation of these elements in AGB stars is given by
\begin{equation}\label{eq:9}
    m_{\mathrm{d}}^\mathrm{AGB} = \begin{cases} \delta^\mathrm{AGB}\left(m_\mathrm{C,ej}^\mathrm{AGB} - 0.75m_\mathrm{O,ej}^\mathrm{AGB}\right) &\mbox{if } m_\mathrm{C,ej}^\mathrm{AGB}/m_\mathrm{O,ej}^\mathrm{AGB} > 1, \\ 
    \\
    \delta^\mathrm{AGB}\left(m_\mathrm{C,ej}^\mathrm{AGB}  + m_\mathrm{N,ej}^\mathrm{AGB} + m_\mathrm{O,ej}^\mathrm{AGB}\right) &\mbox{if } m_\mathrm{C,ej}^\mathrm{AGB}/m_\mathrm{O,ej}^\mathrm{AGB} < 1, \end{cases}
\end{equation}
where $m_\mathrm{d}^\mathrm{AGB}$ is the amount of dust of the $j^\mathrm{th}$ element; $\delta^\mathrm{AGB}$ is the condensation efficiency for AGB stars; $m_\mathrm{C,ej}^\mathrm{AGB}$, $m_\mathrm{N,ej}^\mathrm{AGB}$ and $m_\mathrm{O,ej}^\mathrm{AGB}$ are the mass of carbon, nitrogen and oxygen returned to the ISM by an AGB event, respectively. $\delta_\mathrm{C}^\mathrm{AGB}$ could be treated as free parameter, although for simplicity we adopt the value $\delta_\mathrm{C}^\mathrm{AGB} = 0.2$ from \citet{Popping17}.

For SN II, we track the abundances of C, O, Mg, Si, S, Ca and Fe. The total mass of dust formed in the SN II ejecta, $m_{\mathrm{d}}^\mathrm{SNII}$, is described by
\begin{equation} \label{eq:10}
     m_{\mathrm{d}}^\mathrm{SNII} = \delta^\mathrm{SNII} \left(m_\mathrm{C,ej}^\mathrm{SNII} + m_\mathrm{O,ej}^\mathrm{SNII} + 16 \sum_{j=\mathrm{Mg,Si,S,Ca,Fe}} m_{j,\mathrm{ej}}^\mathrm{SNII} / \mu_j \right),
\end{equation} 
where $m_{j,\mathrm{ej}}^\mathrm{SNII}$ is the mass of $j$th element in the ejecta of SNII and $\mu_j$ is the mass of element $j$ in atomic mass units. $\delta^\mathrm{SNII}$ is the condensation efficiency in SNII and we adopt the value of 0.15 from \citet{Popping17}. This choice accounts for the efficiency with which dust survives the reverse shock \citep{Micelotta16}, and the injection into the ISM.

The total dust mass formed from the stellar ejecta is given by
\begin{equation}
    \dt{M}_\mathrm{d}^\mathrm{formation} = \frac{\mathrm{d}}{\mathrm{dt}}\left(m_\mathrm{d}^\mathrm{AGB}\right) + \frac{\mathrm{d}}{\mathrm{dt}}\left(m_\mathrm{d}^\mathrm{SNII}\right).
\end{equation}

\subsection{Dust growth in molecular clouds}

The existing dust grains grow by accreting refractory elements in dense molecular clouds. This accretion is the second channel for dust production in our model \citep{ZGT08}. To model this process, we follow the procedure presented in \citet{Dwek1998}. The growth rate of dust is described by

\begin{equation} \label{eq:12}
    \dt{M}_\mathrm{d}^\mathrm{growth} = \left(1 - \frac{M_\mathrm{d}}{M_\mathrm{metal}} \right) \left( \frac{f_\mathrm{H_2} M_\mathrm{d}}{\tau_\mathrm{acc}} \right),
\end{equation}
where $M_\mathrm{d}$ is the existing dust mass, $M_\mathrm{metal}$ is the mass of refractory elements, $f_\mathrm{H_2}$ is the fraction of $\mathrm{H_2}$ in the ISM and $\tau_\mathrm{acc}$ is the timescale for dust growth via accretion.

The equation for the accretion timescale is adopted from \citet{Asano13}, defined as

\begin{equation}
\begin{split}
    \tau_\mathrm{acc} &\approx 2.0 \times 10^7 \times \left( \frac{\bar{a}}{0.1\mu \mathrm{m}}\right) \left(\frac{n_\mathrm{H}}{100 \mathrm{cm}^{-3}} \right)^{-1} \left(\frac{T}{50 \mathrm{K}} \right)^{-\frac{1}{2}} \left(\frac{Z}{0.02} \right)^{-1}[\mathrm{yr}] \\
    &= \tau_{\mathrm{acc,0}} Z^{-1}.
\end{split}
\end{equation}
Here $\bar{a}$ is the typical size of grains, $n_\mathrm{H}$ is the volume density of molecular clouds, Z is the abundance of metals and T is the temperature of the molecular clouds. $\tau_{\mathrm{acc,0}} = 4.0 \times 10^5$ yr as we assume $\bar{a} = 0.1 \mu$m, $n_\mathrm{H} = 100 \mathrm{cm}^3$ and T = 50K. 

\subsection{Destruction by supernovae}
Dust in the ISM is mainly destroyed by SN shocks. While this destruction process is complicated, we follow \citet{Asano13} who use the simple analytic prescription presented in \citet{Mckee89} and \citet{DS80}. The dust destruction timescale in this prescription is defined as
\begin{equation}
\label{eq:destruction_timescale}
    \tau_\mathrm{destruct} = \frac{M_\mathrm{ISM}}{f_\mathrm{SN}M_\mathrm{swept}R_\mathrm{SN}},
\end{equation}
where $M_\mathrm{ISM}$ is the total mass of cold gas in the ISM, $f_\mathrm{SN}$ is the efficiency of dust destruction by SN (defined as the ratio of the destroyed dust to the swept dust mass by SN), $M_\mathrm{swept}$ is the total ISM mass swept up by a SN event and $R_\mathrm{SN}$ is the SN rate. We follow \citet{Asano13} to adopt $f_\mathrm{SN} = 0.1$ from \citet{Mckee89} and \citet{Nozawa06}. According to \citet{Mckee89}, the value of $f_\mathrm{SN}$ also accounts for bubbles created by SN that lessen the efficiency of the next SN occurred in that bubble.

The total ISM mass swept up by a SN event depends on the gas-phase metallicity. As metals can be an efficient cooling channel in the ISM, higher gas-phase metallicity will result in smaller swept mass. To compute $M_\mathrm{swept}$ we again follow \citet{Asano13} who use the fitting formula of \citet{Yamasawa11}:
\begin{equation}
    M_\mathrm{swept} = 1535 [(Z/\mathrm{\Zsun}) + 0.039] ^ {-0.289} [\Msun],
\end{equation}
where $\Zsun = 0.02$ is the solar metallicity.

To compute the SN rate $R_\mathrm{SN}$, we follow the definitions in \citet{DC11}:
\begin{equation}
    R_\mathrm{SN} = \frac{\psi}{M_*},
\end{equation}
where $\psi$ is the instantanous star formation rate. $M_*$ is the mass of all stars that were born per SN event, given by

\begin{equation}
    M_* \equiv \frac{\int_{M_l}^{M_u} M \phi(M) dM}{\int_{M_w}^{M_u} \phi(M) dM},
\end{equation}
where $M_l$ and $M_u$ are the lower and upper mass limits of the IMF and $M_w$ is the lower limit of SN stars. Consistent with our chemical enrichment model, we use $M_l = 0.1 M_\odot$, $M_u = 40 M_\odot$ and $M_w = 8 M_\odot$. For the IMF, we use \citet{Chabrier03}.

The rate of dust destruction is then given by
\begin{equation}
    \dt{M}_\mathrm{d}^\mathrm{destruct} = \frac{M_\mathrm{d}}{\tau_\mathrm{destruct}}.
    \label{eq:destruction}
\end{equation}

\subsection{Other physical processes}

Aside from the processes mentioned above that specifically govern the dust content of the ISM, we also include additional processes that change the dust contents via mass transfer from/to the entire ISM. We follow \citet{Popping17} to incorporate these processes.
\subsubsection{Dust locked up in stars}
Stars form in molecular clouds in the ISM. When the ISM is polluted with metals and dust, the newly born stars also contain a fraction of dust and metals. The rate of dust captured in every star formation episode is proportional to the star formation rate $\psi$ and the dust-to-gas (DTG) ratio $\frac{M_\mathrm{d}}{M_\mathrm{ISM}}$, described as
\begin{equation}
    \dt{M}_\mathrm{d}^\mathrm{SF} = \frac{M_\mathrm{d}}{M_\mathrm{ISM}}\psi.
\end{equation}

\subsubsection{Inflows and outflows}
The energy from SN and AGN can heat up and eject material out of the ISM; this process is called feedback. In every outflow, we assume that the DTG ratio of the ejected gas equals the DTG ratio of the ISM. Hence, the rate at which dust is ejected from the ISM is proportional to the total outflow rate of the ISM ($\dt{M}_\mathrm{ISM}^\mathrm{out}$) and DTG ratio, expressed as
\begin{equation}
    \dt{M}_\mathrm{d}^\mathrm{out} = \frac{M_\mathrm{d}}{M_\mathrm{ISM}} \dt{M}_\mathrm{ISM}^\mathrm{out}.
\end{equation}

The mass expelled out of the ISM can be re-accreted into the ISM later through inflow. The balance between inflow and outflow is critical in regulating the overall properties of the ISM. Similar to the outflow process, the dust inflow rate is described by
\begin{equation}
    \dt{M}_\mathrm{d}^\mathrm{inflow} = \frac{M_\mathrm{d}^\mathrm{hot}}{M_\mathrm{gas}^\mathrm{hot}}\dt{M}_\mathrm{gas}^\mathrm{inflow},
\end{equation}
with $\dt{M}_\mathrm{gas}^\mathrm{inflow}$ the total inflow rate of the gas, $M_\mathrm{d}^\mathrm{hot}$ is the total dust mass in the hot halo and $M_\mathrm{gas}^\mathrm{hot}$ is the total gas mass in the hot halo. We assume that the DTG ratio of the reaccreted mass equals the DTG ratio of the hot gas.

\subsubsection{Mergers}
Mergers between central and satellite galaxies include mass transfer that affects the morphology of the central galaxy. In the original merger prescription of \texttt{SAGE} \citep{Croton06, Croton16}, the output of a merger depends on the mass ratio of the merging galaxies. If the mass ratio is smaller than 0.3, a minor merger occurred. In a minor merger, the stellar content of the satellite is added to the central's bulge while the ISM content (cold gas, metals and dust) is added to the central's disk along with the stars formed during a minor merger starburst. On the other hand, if the mass ratio of the progenitors exceeds 0.3, a major merger occurred where the starburst activity is significantly higher than a minor merger. In a major merger, the disks of both galaxies are destroyed and all stars are placed in a spheroid. When a merger happened in \texttt{Dusty SAGE} we assume that the dust undergoes the identical process as the cold gas but scaled by the DTG ratio. No grain destruction is assumed in the merger.

\subsection{Thermal sputtering in the hot halo}
The environment in the hot halo and the ejected reservoir can destroy dust grains via thermal sputtering. We follow \citet{Popping17} to incorporate a thermal sputtering prescription. For simplicity, we assume that the density and temperature in the ejected reservoir equals that in the hot halo. 

The rate of sputtering in the hot halo and the ejected reservoir is defined as
\begin{equation}
    \frac{da}{dt} = -(3.2 \times 10^{-18}) \left(\frac{\rho}{m_p} \right) \left[\left(\frac{T_0}{T} \right)^{\omega} + 1 \right]^{-1} [\mathrm{cm} \mathrm{s}^{-1}],
\end{equation}
where $a$ is the grain radius, $T$ and $\rho$ are the temperature and density of the hot gas, $m_p$ is the proton mass, $T_0 = 2 \times 10^{-6} \mathrm{K}$ is the critical temperature above which the sputtering rate flattens, and $\omega = 2.5$ controls the sputtering rate in the low-temperature regime. The initial size of dust grains before sputtering is taken to be $0.1 \mu \mathrm{m}$. The sputtering time scale of the grain is then described as \citep{TM95}
\begin{equation}
    \tau_\mathrm{sputtering} = 0.17 \left(\frac{a}{ 0.1 \mu \mathrm{m}}\right) \left(\frac{\rho}{10^{-27} \mathrm{g cm}^{-3}} \right)^{-1} \left[\left(\frac{T_0}{T} \right)^\omega + 1 \right] [\mathrm{Gyr}].
\end{equation}

We assume that the temperature $T$ in both hot halo and ejected reservoir reach the virial value. The destruction rate for hot dust due to thermal sputtering is described in \citet{McKinnon16} as
\begin{equation}
    \dt{M}_\mathrm{d}^\mathrm{sputtering} = -\frac{M_\mathrm{d}}{\tau_\mathrm{sputtering} / 3}.
\end{equation}
The amount of dust destroyed by sputtering is added back to the metals in the hot halo.

When computing the cooling rate for the gas in the hot halo, we follow the original recipe from \citet{Croton16, Croton06} and add dust as one of the cooling channels. For now, we use the same treatment for dust and metals in the cooling recipe, but we plan to include a more detailed dust cooling process \citep[e.g.][]{Vogelsberger19, OS73} in the future.

\section{Results and Discussions}
\label{sec:result}

We run our model, \texttt{Dusty SAGE}, on the Millennium simulation \citep{Springel05}. This cosmological N-body simulation evolves $2160^3$ particles in a box of length $500 h^{-1} \mathrm{Mpc}$ from redshift $z=127$ to $z=0$. Dark matter halos are found using the \texttt{SUBFIND} algorithm \citep{2001Springel}, and each consist of at least 20 particles, with a particle mass being $8.6 \times 10^8 h^{-1}$ \Msun. Halo evolution is tracked using the consistent trees code, from which the simulation merger tree files are produced; these serve as the input to \texttt{Dusty SAGE}.

\subsection{Observational datasets}
To constrain \texttt{Dusty SAGE}, we use the observed dust mass in galaxies inferred from different techniques. We summarize this data below, which features throughout the rest of this work, and each measuring technique.

The cold dust content in galaxies is measured from its infrared emission. Several SED models have been developed to infer the dust content from the infrared and sub-mm observations. The most straightforward method is by fitting a single modified black body (MBB) curve to the SED. \citet{Mancini15} apply this method to ALMA and PdBI observations to measure an upper limit of the dust mass in five galaxies at $z=6$ and four galaxies at $z=7$. \citet{Eales09} also apply a single MBB to data obtained from the Balloon-borne Large Aperture Submillimeter Telescope (BLAST), converting its luminosity function into a dust mass function. 

For cases where the dust temperature range can not be described using a single value, it is necessary to use a more complicated prescription. \citet{VDE05} use a double temperature source assumption to infer the dust mass function from the galaxies in the SCUBA (Submillimetre Common-User Bolometer Array) Local Galaxy Survey (SLUGS) and the IRAS Point Source Catalog Redshift Survey (PSCz). \citet{DEE03} also use two MBBs to construct the dust mass function in the local and high redshift Universe using samples from SLUGS and deep SCUBA submillimetre observations.

\texttt{MAGPHYS} \citep{daCunha08} is a SED fitting tool that applies an energy balance technique to interpret the mid and far infrared emission of galaxies. \citet{daCunha15} used it to interpret the dust properties of 122 sub-mm galaxies from the ALMA LESS survey. \citet{Clemens13} also used \texttt{MAGPHYS} to interpret the dust properties of local star-forming galaxies from \textit{Planck}, WISE, \textit{Spitzer}, IRAS and \textit{Herschel} combined data.

Beyond this, several authors have layered additional complexity in their modelling to interpret the dust emission more accurately, accounting for the distribution of a radiation field heating the dust and the composition of dust grains. \citet{DL07} used a mixture of amorphous silicate, graphite and varying amounts of PAHs to model the infrared emission. \citet{Santini14} use this model to measure the dust mass of galaxies in the GOODS-S, GOODS-N and COSMOS fields, observed using \textit{Herschel}, PACS and SPIRE. \citet{Ciesla14} model galaxies in the \textit{Herschel} Reference Survey in a similar way. \citet{Galliano11} model the dust SED with two plausible grain compositions: the standard composition consists of PAH, silicates and graphite (hereafter \textit{gr}) and the alternate consists of PAH, silicates and amorphous carbon (hereafter \textit{ac}). \citet{RR14} applied this model to two samples of local galaxies from the \textit{Herschel} Dwarf Galaxy Survey (DGS) and the Key Insight on Nearby Galaxies: a Far-Infrared Survey with Herschel (KINGFISH). They find that the dust mass obtained from the alternate composition is lower than the standard graphite composition. \citet{Jones16} developed the \texttt{THEMIS} dust model using the properties of silicate-core carbonaceous-mantle grains measured in the laboratory. \citet{Nersesian19} implement \texttt{THEMIS} into the \texttt{CIGALE} SED fitting code \citep{BBI05} to infer the properties of 875 nearby galaxies measured with \textit{Herschel} and published the results in the DustPedia archive \citep{Davies17}.

In \texttt{Dusty SAGE}, we use the dust mass function at $z=0$ \citep{VDE05, Clemens13, Dunne11} and the dust mass - stellar mass relation at $z=0$ \citep{RR14, Santini14, Nersesian19} as our primary model constraints.

\subsection{Evolution of dust in the ISM}

\begin{figure}
    \centering
    \includegraphics[width=0.5\textwidth]{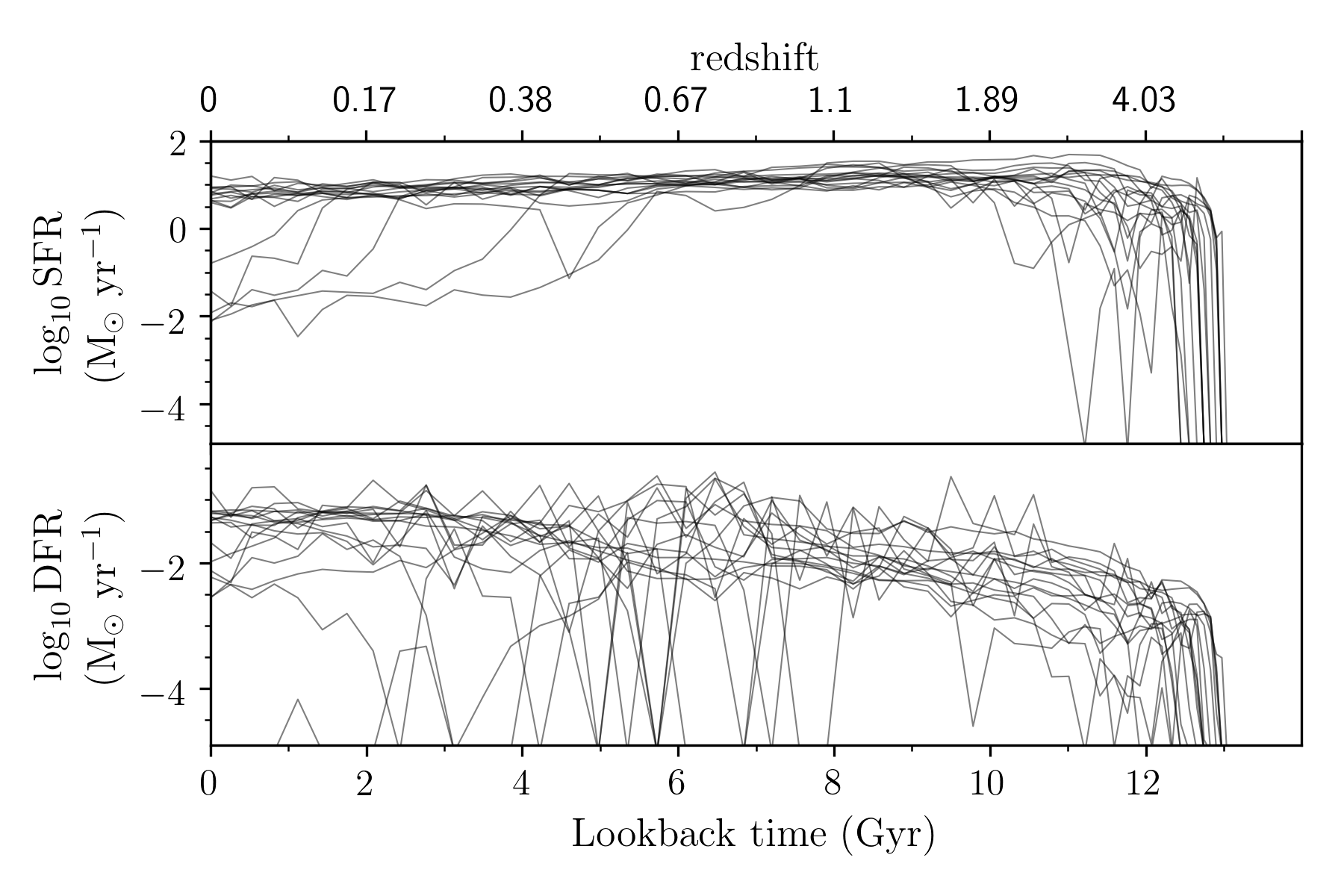}
    \caption{\textit{Top panel} shows the star formation rate and \textit{bottom panel} shows the net dust formation rate of Milky Way-type model galaxies in \texttt{Dusty SAGE}.}
    \label{fig:SFH-DFH}
\end{figure}

In this section, we use our model to predict the evolution of the dust content in galaxies from redshift $z = 7$ to redshift $z = 0$. We select all \texttt{Dusty SAGE} galaxies with stellar mass $M > 10^8$ $\Msun$ and plot several dust scaling relations, including the dust mass - stellar mass relation, DTG ratio, DTM ratio and dust mass function. 

Figure \ref{fig:SFH-DFH} shows the evolution of the dust and star formation rate of Milky Way-type galaxies in \texttt{Dusty SAGE}. We select galaxies based on their stellar mass ($M_\mathrm{star} = 5.8 \times 10^{10} - 1 \times 10^{11} \Msun$, morphology (bulge to stellar mass ratio = 0.35 - 0.4), and star formation activity (SFR > 0 \Msun $\mathrm{yr}^{-1}$). The net dust formation rate (DFR) is $\sim 2$ dex lower than the SFR. Although the majority of the galaxies follow a general trend, both in the star formation and dust formation history, we can see that for some galaxies SFR starts declining earlier than their counterparts. Considering their dust formation, many galaxies have DFR = 0 during their history, indicating that the destruction rate at that time overcame the total production rate.

\subsubsection{Dust mass function}
\label{sssec:dmf}
\begin{figure*}
    \centering
    \includegraphics[width = 1.0\textwidth]{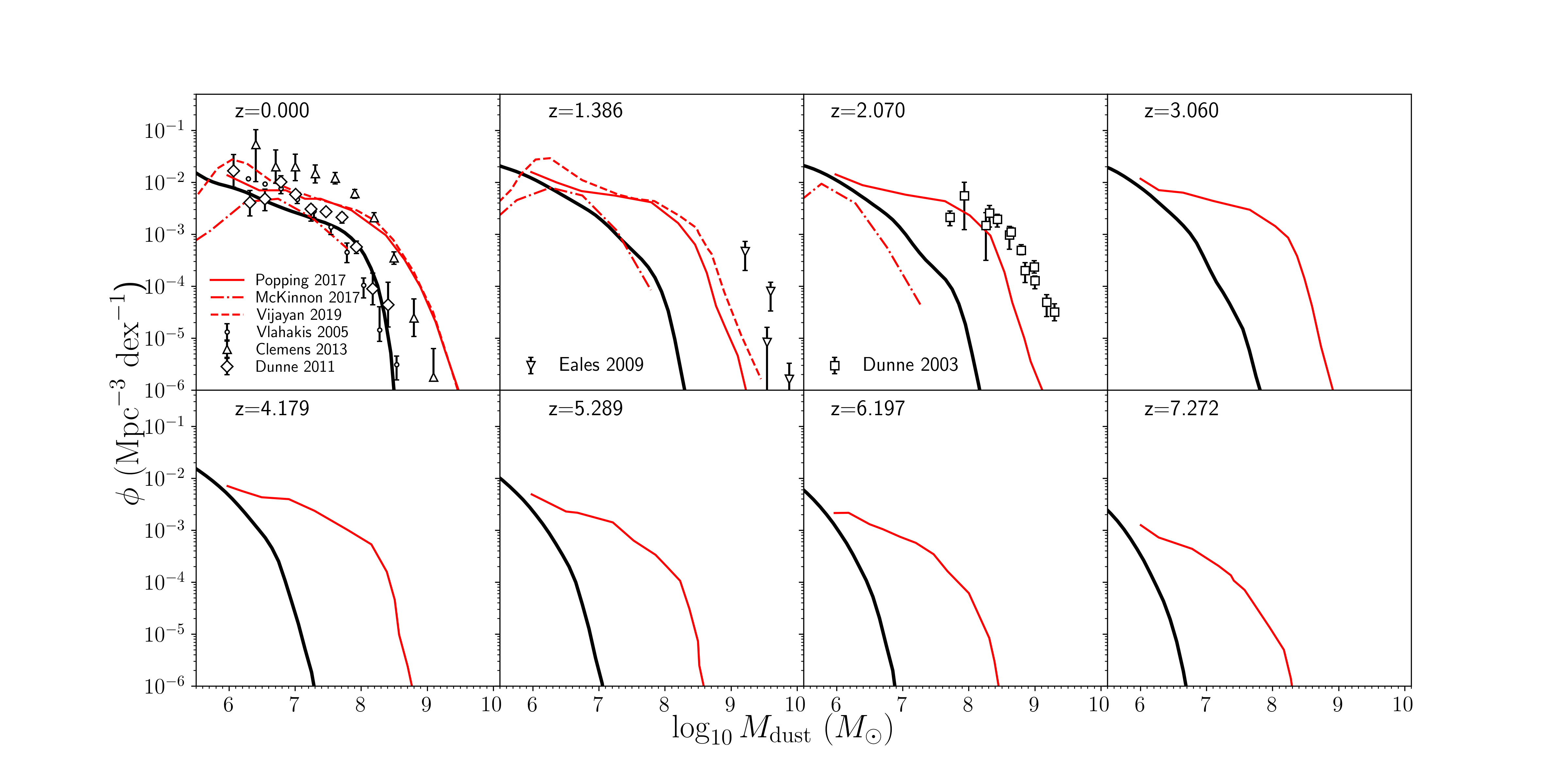}
    \caption{The evolution of the dust mass function from redshift $z=0$ to redshift $z=7$. Our predictions are marked with black solid lines while the red solid, red dash-dotted, and red dashed lines are the predictions from \citet{Popping17}, \citet{McKinnon17}, and \citet{Vijayan19}, respectively. The models are compared with observations from \citet{DEE03}; \citet{VDE05}; \citet{Eales09}; \citet{Dunne11} and \citet{Clemens13}.}
    \label{fig:11}
\end{figure*}

We present the evolution of our predicted dust mass function between redshift $z=7$ and the present day in Figure \ref{fig:11}. At redshift $z=0$, we compare with the observations of \citet{Dunne11}, \citet{VDE05}, and \citet{Clemens13}. Our dust mass function follows a Schechter function with the knee at a dust mass $\sim 10^8 \Msun$. This is in good agreement with \citet{Dunne11}. It also provides a good fit to \citet{VDE05} for galaxies with low dust mass, but shows a higher number density around the knee. Compared to \citet{Clemens13}, our model underpredicts the number density across the entire mass range. 

At redshift $z=1$, we compare our predicted number density with the observational dataset at redshift $z=0.6$ to redshift $z=1$ from \citet{Eales09}. At redshift $z=2$, we use the high redshift sample in \citet{DEE03}. These high redshift observations mostly populate the high mass end of the mass function, and our model provides a poor fit. However, we note that the observations are based on submillimetre surveys with a large source beam. Arcsecond-resolution observations with ALMA report that the brightest sources in such surveys are likely comprised of multiple fainter sources \citep{Karim13}.

Our predictions show a clear evolution in the dust mass function from the present day to redshift $z=7$, where a decrease in amplitude can be seen above redshift $z=3$.  Between redshift $z=3$ and redshift $z=0$, the number density is roughly constant for galaxies with the low dust mass but increases for galaxies with high dust mass. On the other hand, the observations reveal an opposite trend. For example, \citet{DEE03} showed that the number density of galaxies with high dust mass is lower in local galaxies than for their high redshift sample. But again, it is possible that their high redshift sources consists of multiple galaxies.

We compare our results with predictions from \citet{Popping17}, \citet{Vijayan19}, and \citet{McKinnon17}. The evolution of the dust mass function from \citet{Popping17} shows the similar trend of increasing over time. Both \citet{Popping17} and \citet{Vijayan19} produce a Schechter function with the knee at higher dust mass than our prediction at all redshifts. Conversely, results from \citet{McKinnon17} give a good match with our ISM dust at redshift $z=0$ and $z=1$. We can see that all of the models considered are in a good agreement with the broad range of data locally. However, no model appears to match the observations at redshift $z > 1$. Note that each model shows the dust mass function for the cold dust mass in the ISM only, while the dust temperature from the observations at high redshift may be uncertain. Further failures in reproducing galaxy properties driven by the uncertainty of the underlying galaxy modelling at high redshift can also affect dust production mechanism. To match the observed dust mass function at $z>1$, we might need a prescription for dust production and destruction that allows more variation with galaxy properties, therefore evolving with time. For example, dust production in the ejecta of SN may be higher at early times \citep{Dwek14}, or destruction rate varies with the ambient gas density and metallicity \citep{Temim15}.

\subsubsection{Dust mass vs stellar mass}
\label{sssec:dust-stellar mass}
\begin{figure*} 
    \centering
    \includegraphics[width = 1.0\textwidth]{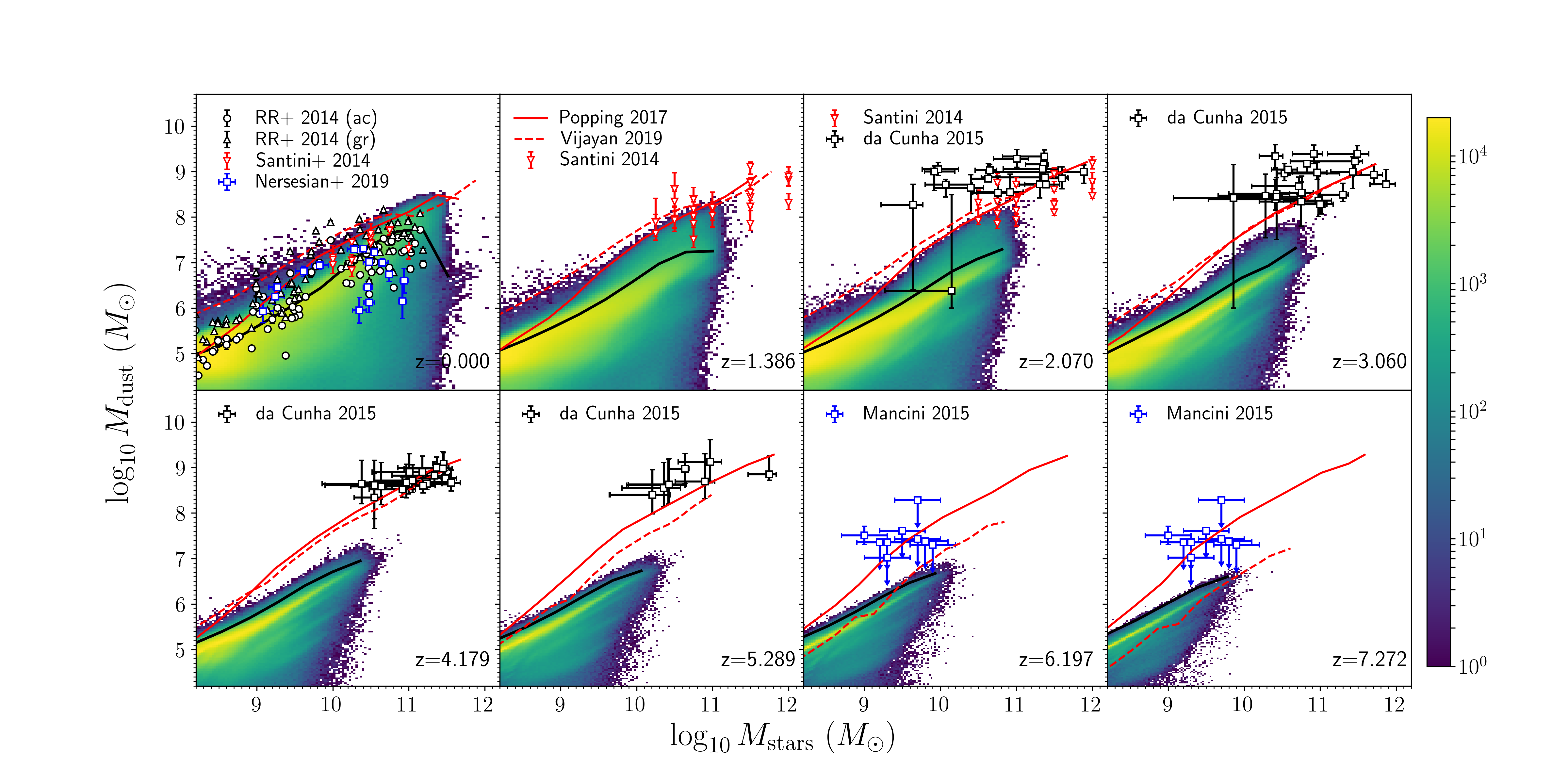}
    \caption{The dust mass of galaxies as a function of stellar mass from redshift $z=0$ to redshift $z=7$. The heat map shows the 2D density distribution of galaxies in our model with brighter color representing higher density, while black lines mark the median. The red solid and red dashed lines are the predictions from \citet{Popping17} and \citet{Vijayan19}, respectively. The symbols with errorbar are the observational values from \citet{RR14, Nersesian19, Santini14, daCunha15, Mancini15}, as indicated in the legend.}
    \label{fig:3} 
\end{figure*}

Observations have shown a relation between dust mass and stellar mass across an extensive redshift range. We present the evolution of this relation from our model in Figure \ref{fig:3}. At all redshifts, the dust masses increase linearly with stellar mass. However, at fixed stellar mass, there is a large scatter of dust masses under the median, revealing a dust-poor galaxy population across all stellar masses. The upper envelope of the relation is remarkably tight, marking a maximum dust mass, which is $\sim 1\%$ of their stellar mass. In our prescription, dust can no longer grow once all metals have turned into dust, hence the tight upper limit. 

We can see that at $z=0$, our model is in good agreement with the observations across the entire mass range. This includes the relation found in \citet{RR14}, although our model matches the dust mass obtained with amorphous carbons composition better than the graphite composition. This is aligned with studies that find that the carbonaceous dust in the ISM is more likely to exist in the amorphous form than in graphitic form. For example, \citet{SDJ08} investigate the carbonaceous grain material in the ISM and find that the amorphous carbon is more efficiently destroyed by shocks than graphite, leading to a better agreement with the observations taken from shocked regions. They concluded that even if a lot of graphite grains are injected into the ISM, they are unlikely to remain graphitic due to erosion and ion irradiation. For the remainder of this work, we refer to the dust mass derived by a model with amorphous carbon composition when considering results from \citet{RR14}.

The median value predicted by our model agrees with the relation found in \citet{Santini14} for stellar mass $\log M < 11$ \Msun. Above this point, our median value is $\sim 0.5$ dex too large. Our model is consistent with the DustPedia data \citep{Nersesian19} up to stellar mass $\log M = 10$ \Msun. Above this point, our median dust mass is $\sim 1$ dex too large, but the observations still lie within the scatter in our plot.

We can see that the normalization of this relation decreases $\sim 0.4$ dex from redshift $z=7$ to $z=0$, highlighting how dust formation began in the early Universe. The observations from \citet{Santini14} at redshift $z = 1$ and $z = 2$ lies $\sim 0.5$ dex higher than our median value for stellar mass $\log M = 11$ \Msun and we do not produce galaxies with stellar mass $\log M > 11$ \Msun. Our model predicts a dust mass that is one order of magnitude lower than the observations from \citet{daCunha15} for galaxies with stellar mass $\log M = 10.5$ \Msun from redshift $z = 2$ to $z = 4$, and $\log M = 10$ \Msun at redshift $z=5$. Again, our model fails to reproduce galaxies with stellar mass $\log M > 10.5$ \Msun at redshift $2-4$ and $\log M > 10$ \Msun at redshift $z=5$. However, we should note that the \citet{daCunha15} sample is dominated by the extremely brightest galaxies at each particular redshift. At redshift $z = 6$ and $z = 7$, our model shows a $\sim 0.5$ dex offset from the upper limit in \citet{Mancini15}.

We also plot the results from \citet{Popping17} and \citet{Vijayan19} in Figure \ref{fig:3}. \citet{Popping17} have galaxies with stellar masses up to $\log M = 11$ \Msun at all redshifts and a steeper slope than our model, allowing them to provide a better agreement to high redshift observation. The steeper slope might be due to their high dust production rate density compare to our model. Results from \citet{Vijayan19} show higher normalizations than our model at redshift $z < 4$ but lower at redshift $z > 4$. Their model also does not produce the most massive galaxies at high redshift due to the limitation of their simulation box size. Still, they show galaxies with stellar mass higher than ours at high redshift, using the newest version of \texttt{L-Galaxies} SAM \citep{Henriques15} calibrated to produce such up to redshift $z=3$. Our base model, \texttt{SAGE} \citep{Croton06, Croton16}, is constrained using a broad set of galaxy observation at redshift $z=0$ only.

\subsubsection{Dust-to-gas (DTG) ratio}
\label{sssec:dtg}
\begin{figure*}
    \centering
    \includegraphics[width = 1.0\textwidth]{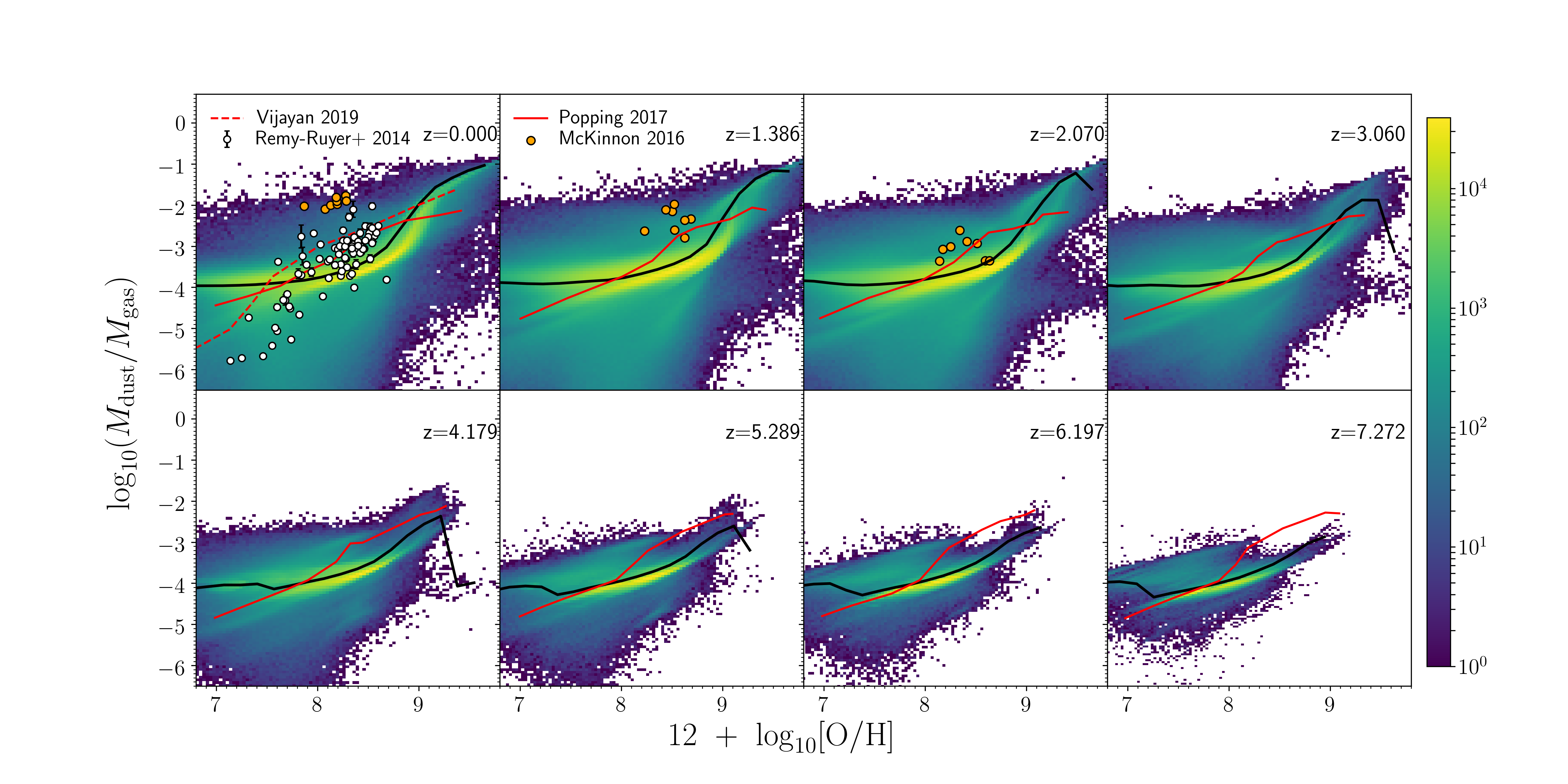}
    \caption{The dust-to-gas ratio as a function of gas phase metallicity from redshift $z=0$ to $z=7$. The heat map shows the 2D density distribution of galaxies in our model with brighter color representing higher density, while black lines mark the median. The red solid and red dashed lines are the predictions from \citet{Popping17} and \citet{Vijayan19}, respectively. The orange circles are the prediction for eight Milky-Way sized galaxies from \citet{McKinnon16}. Our predictions at redshift $z=0$ are compared to the observations by \citet{RR14}.}
    \label{fig:5}
\end{figure*}

We present the evolution of the relationship between the DTG ratio and metallicity in the form of oxygen abundance in Figure \ref{fig:5}. At all redshifts, the DTG ratio increases with the gas phase metallicity. The rise is slow for galaxies with low metallicity and more rapid for galaxies with oxygen abundance $> 8.5$, before it reaches a turning point around $12 + \log[\mathrm{O/H}] \sim 9.2$. Above this point, the slope becomes shallow again. 

At redshift $z=0$, we compare the DTG ratio of our model galaxies to the observational constraint from \citet{RR14}. We derive the DTG ratio in \citet{RR14} using their amorphous carbon dust mass. First we compute the total cold gas mass in each galaxy in the sample using Equation 2 in \citet{RR14}. Then we divide their amorphous dust mass with the computed gas mass to get the DTG ratio. The metallicities in \citet{RR14} are derived using strong emission line methods and the calibration from \citet{Pilyugin05}. They assumed a global metallicity value for each galaxy. We find a good agreement with the observations for galaxies with oxygen abundance $\sim 8.5$. Below this point, the median DTG ratio of our predicted galaxies is higher than the observations, although we do find model galaxies in this regime that represent the observed galaxies.

At higher redshifts, especially at redshift $z > 4$, the normalization of the DTG relation clearly decreases with lookback time. The gradient of the relation is relatively stable from $z=7$ to $z=4$, but then steepens between redshifts $z=3$ and $z=0$. This steepening occurs in the redshift range where the dust growth dominates the dust production mechanism. We will revisit the effects of the dominant dust production mechanism to DTG ratios in Section \ref{sssec:relation}.

The predictions from \citet{Popping17} and \citet{Vijayan19} roughly matches our prediction for oxygen abundance $> 8$ at all redshifts. For galaxies with low metallicity, their prediction is one order of magnitude smaller than our median value. The relation in \citet{Vijayan19} shows a steepening at around $12 + \log[\mathrm{O/H}] \sim 7.2$ which becomes shallower again around $12 + \log[\mathrm{O/H}] \sim 8.0$. This behaviour is not seen from the \citet{Popping17} model. The steeper slope in this relation is caused by an efficient grain growth mechanism while the shallower slope occur when stellar dust production dominates. Our model and \citet{Vijayan19} (their Figure 8) shows that there is a switch in dust production rate evolution, while Figure 10 in \citet{Popping17} shows that their growth rate exceeds the stellar production rate at all redshifts. The DTG ratio predicted by \citet{McKinnon16} is $2$ dex too high at redshift $z=0$ and about $0.5$ dex too high at $z=1$, indicating a high metal depletion at low redshift. Their model is in good agreement with ours at redshift $z=2$. 

\begin{figure*}
    \centering
    \includegraphics[width = 1.0\textwidth]{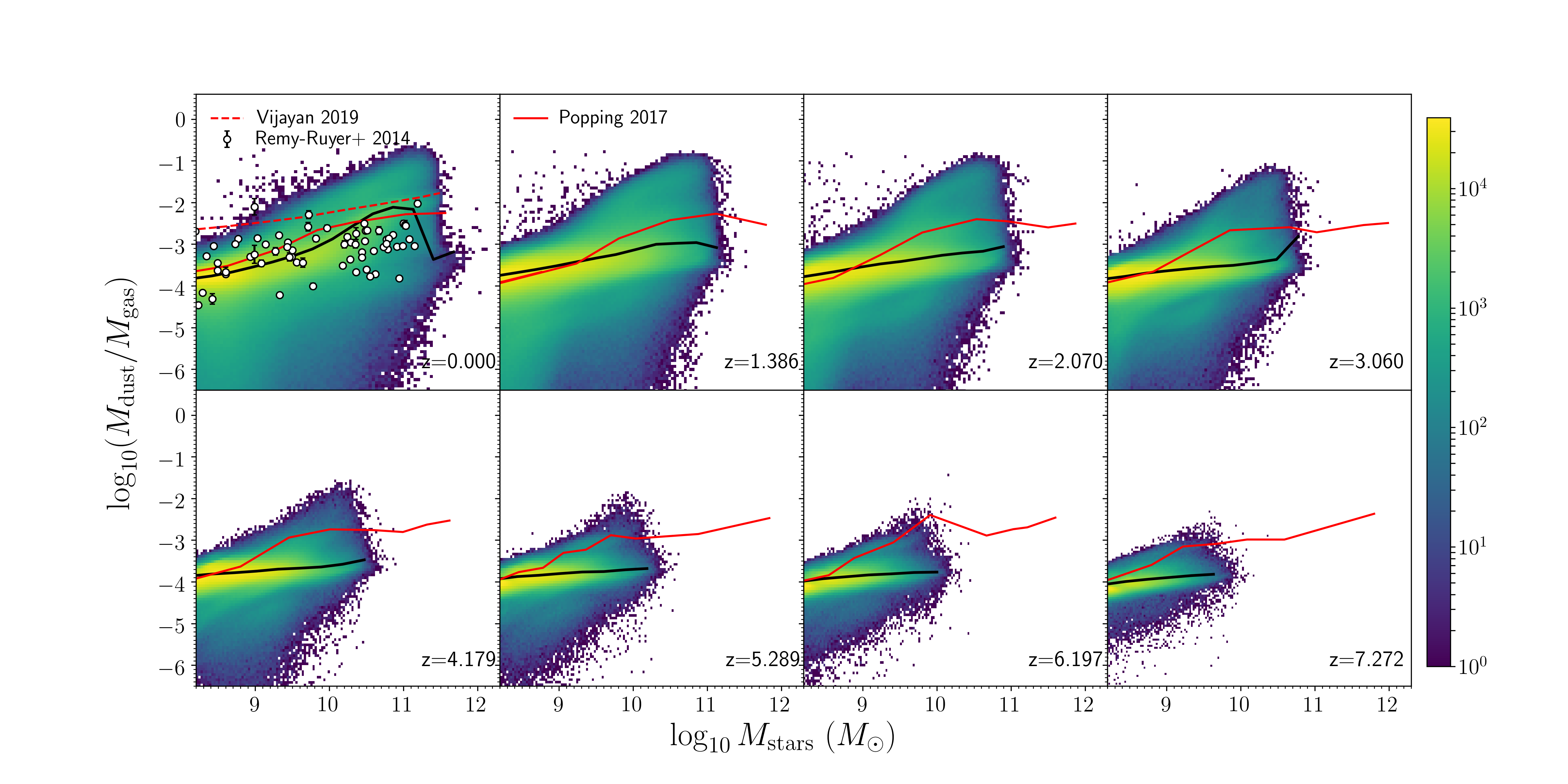}
    \caption{The dust-to-gas ratio as a function of stellar mass from redshift $z=0$ to $z=7$. The heat map shows the 2D density distribution of galaxies in our model with brighter color representing higher density, while black lines mark the median. The red solid and red dashed lines are the predictions from \citet{Popping17} and \citet{Vijayan19}, respectively. Our predictions at redshift $z=0$ are compared to the observations by \citet{RR14}.}
    \label{fig:6}
\end{figure*}

We plot the relation between DTG ratio and stellar mass from redshift $z=7$ to $z=0$ in Figure \ref{fig:6}. The DTG ratio increases with stellar mass, though the slope is not as steep as the relation with metallicity. At redshift $z=0$, we again compare our prediction with the observations from \citet{RR14} and find that our predictions overlap with the data up to a stellar mass $M_* = 10^{10} \Msun$. Above this, our median DTG is slightly higher than the observations. Considering the evolution of the DTG ratio vs stellar mass relation, we can see a gradual decrease in the gradient from the present day to redshift $z=7$, and more significantly around redshift $z=3$. Again, this is the regime where dust growth in molecular clouds starts to dominate (see Figure \ref{fig:DFRD}). The overall relation shifts $\sim 0.2$ dex lower out to redshift $z=7$. Our model is roughly consistent with the results from \citet{Popping17} at all redshifts. Their slope is steeper than ours at redshift $z>1$. Our model also matches the result from \citet{Vijayan19} for galaxies with stellar mass $\log M > 10$ \Msun. Below this mass, our predictions lie $0.5$ dex lower than theirs.

\subsubsection{Dust-to-metal (DTM) ratio}
\label{sssec:dtm}
\begin{figure*}
    \centering
    \includegraphics[width = 1.0\textwidth]{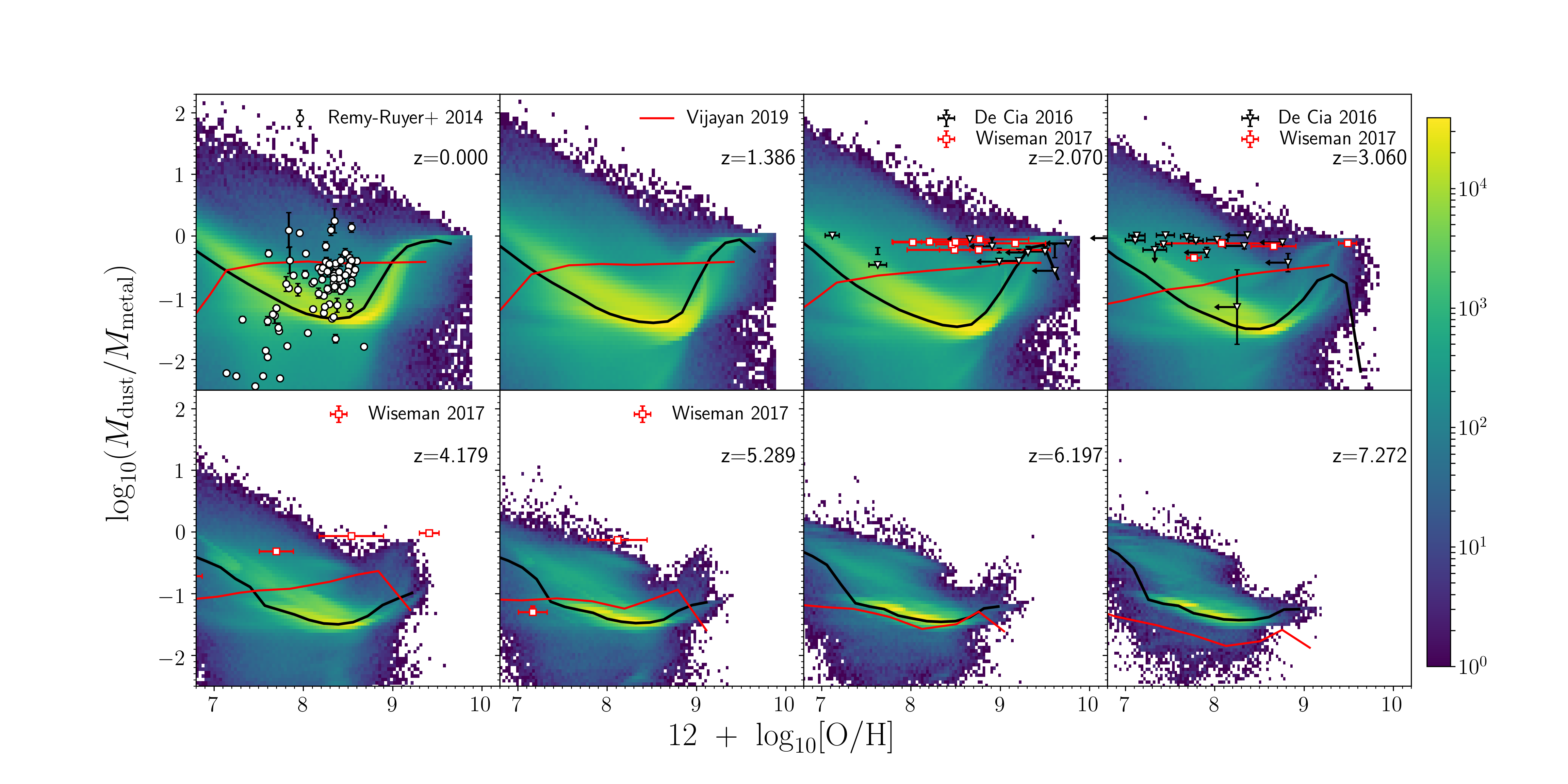}
    \caption{The dust-to-metal ratio as a function of gas phase metallicity from redshift $z=0$ to $z=7$. The heat map shows the 2D density distribution of galaxies in our model with brighter color representing higher density, while black lines mark the median. The red solid lines are the predictions from \citet{Vijayan19}. Our predictions at redshift $z=0$ are compared to the dust-to-metal ratios derived using datasets in \citet{RR14}.}
    \label{fig:8}
\end{figure*}

We present the relation between the dust-to-metal (DTM) ratio and gas phase metallicity in the form of oxygen abundance from redshifts $z=7$ to $z=0$ in Figure \ref{fig:8}. We compare our predictions at redshift $z=0$ with the DTM ratio we derived from data in \citet{RR14}. First, we use the value of $12 + \log [\mathrm{O/H}]$ and gas mass in \citet{RR14} to compute the total metal mass in each sample galaxy, assuming $12 + \log[\mathrm{O/H}] - \log[Z/Z_\odot] = 9$ and a solar metallicity of $0.02$. Then, we divide the dust mass obtained using the amorphous carbon composition with the derived metal mass. Most of the observed DTM ratio overlaps with our model galaxies. However, we notice the large scatter in the DTM ratio that does not quite match our prediction. This scatter could be caused by the large systematic uncertainties when estimating the gas phase metallicities of the ISM. 

At higher redshift, our predictions give a good fit with the observational constraint from damped Lyman-$\alpha$ emitters \citep{DeCia16}, but lower values with the observations from gamma-ray bursts \citep[GRBs,][]{WSB17}. In the GRB observations, no correlation is found between the DTM ratio and metallicity. This is intriguing because in nearby galaxies, the DTM ratio increases with metallicity. A likely explanation is that the method to deduce both dust mass and metallicity using GRBs causes a systematic offset with the values inferred from the stellar and dust emission.

The overall shape of the DTM ratio vs metallicity relation mirrors the DTG ratio relation. At redshift $z<4$ in both relations, we find a sharp change of gradient around an oxygen abundance $\sim 8.5$. Above this point, the relation increases rapidly. These features are caused by a change in the dust production channel, which will be discussed in the next section. 

Figure \ref{fig:8} also shows the evolution of the DTM relation from \citet{Vijayan19}. Although their model does not predict gradient changes in the relation like ours, their median value is consistent with our median value from redshift $z=0$ to $z=6$. At redshift $z=7$, our prediction is $0.3$ dex higher than theirs but both results have similar slope.  

\begin{figure*}
    \centering
    \includegraphics[width = 1.0\textwidth]{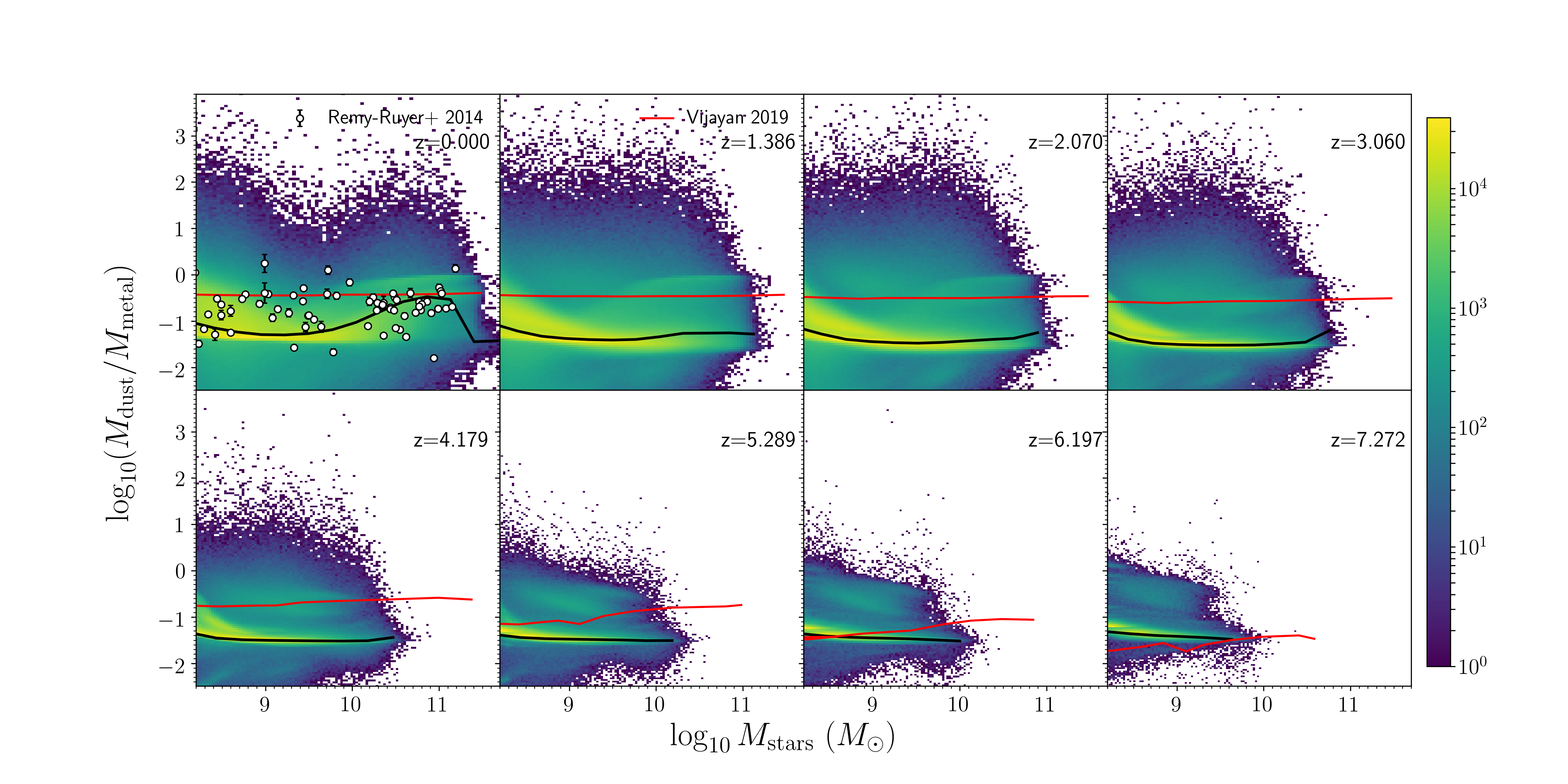}
    \caption{The dust-to-metal ratio as a function of stellar mass from redshift $z=0$ to $z=7$.  The heat map shows the 2D density distribution of galaxies in our model with brighter color representing higher density, while black lines mark the median. The red solid lines are the predictions from \citet{Vijayan19}. Our predictions at redshift $z=0$ are compared to the dust-to-metal ratios derived using datasets in \citet{RR14}.}
    \label{fig:9}
\end{figure*}

We show the evolution of the DTM ratio and stellar mass relation in Figure \ref{fig:9} and we again compare our prediction at redshift $z=0$ with the derived values from \citet{RR14} and find a good agreement. In the evolution of DTM ratio vs stellar mass relation, we see that the relation is stable from redshift $z=7$ to $z=4$ and the slope gradually increases from redshift $z=3$ to $z=0$. This evolution in the trend is similar to the evolution of the DTG ratio vs stellar mass relation. We find a $0.5$ dex offset with the relation in \citet{Vijayan19} from redshift $z=0$ to $z=3$. This offset shrinks at higher redshift, and the models are in rough agreement at redshifts $z=6$ and $z=7$. This shows that the evolution in their model is stronger than ours.

\subsection{The dominant drivers of dust content in galaxies and their influence to DTG and DTM ratios}
\label{ssec:dominant shift}

\begin{figure*}
    \centering
    \includegraphics[width=0.8\textwidth]{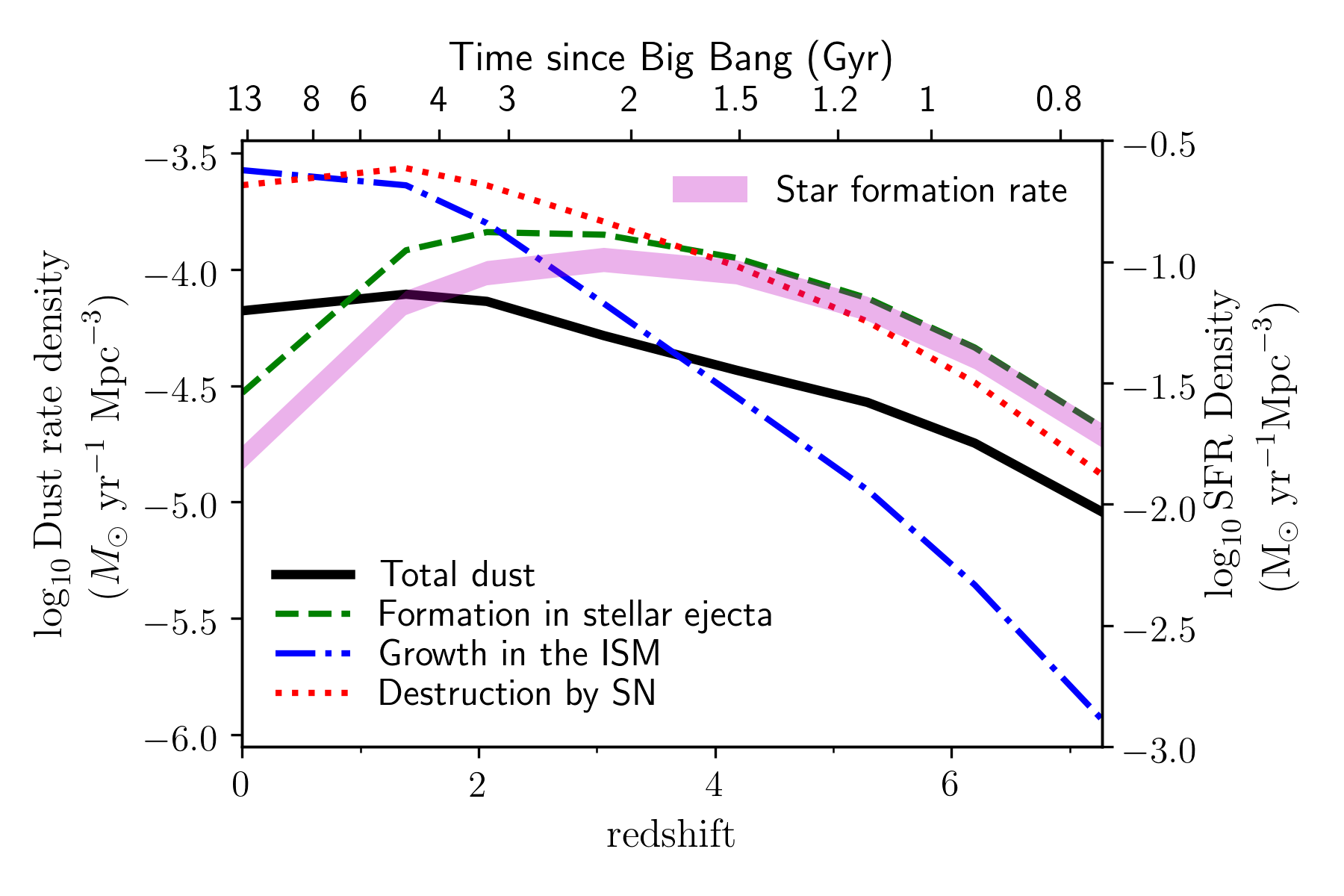}
    \caption{The rate density of different dust production/destruction mechanisms in the ISM as a function of redshift and time since the Big Bang. Thick solid black, green dashed, blue dash-dotted and red dotted lines represent the total rate density, the formation rate density in stellar ejecta, the growth rate density in the ISM, and the destruction rate density by SN, respectively. The magenta shaded regions shows the star formation rate density of our model, with the scale in the right axes.}
    \label{fig:DFRD}
\end{figure*}

In this section, we explore the contribution of each of the primary physical processes that drive the dust content in the ISM at different redshifts. The rate density for each production/destruction mechanism is shown in Figure \ref{fig:DFRD}. This plot shows that the total dust production rate increases sharply towards $z = 2$, reaches its peak between $z=2$ and $z=1$, then decreases towards the present day. The total production rate is the net of all three processes: formation in stellar ejecta, growth via accretion in the ISM and destruction by SN. We can see the same general trend in all three mechanisms; the rate density reaches a peak at redshift $z \sim 1-3$ and then decreases to redshift $z=0$. 

In the early Universe, the dominant mechanism for dust production is its formation in stellar ejecta. Between redshift $2 < z < 3$, the dust growth in molecular clouds overtakes the initial formation. This redshift range is a critical shifting point in the dust production mechanism. After this point, the formation rate in stellar ejecta decreases and the dust growth becomes the main production channel for the overall dust mass. For the overall rate density, the formation in stellar ejecta dominates in the early Universe, from $z>7$ to $z = 4$. Later, between $1 < z < 4$, the destruction process (red dotted line) dominates the overall process. However, the total dust rate density (thick solid black line) in this redshift range still grows because the net formation rate density in stellar ejecta and grain growth exceeds the dust destruction rate density. At about $z \approx 0.5$, the grain growth rate density exceeds the destruction and becomes the main driver for galactic dust content for the present Universe.

Our results imply that the first dust in galaxies is created via the condensation of metals expelled by stars. The existing dust in the ISM then grows by accreting more materials in the dense molecular clouds. The shift to when grain growth in molecular clouds starts to dominate is a critical point that affects the dust properties before and after the pont (see the next section).

The trend for stellar dust production rate density closely resembles the star formation rate density; they increase $\sim 0.5$ dex from redshift $z=7$ to $z \sim 2$, then decrease almost to their original value at present. The similarity is due to the strong dependence of the stellar dust production with the star formation activity. Conversely, the growth rate density steeply increases by more than $2$ dex from redshift $z=7$ to $z \sim 2$, then flattens to the present time. The global destruction rate closely follows production in stellar ejecta for $z > 2$, then follows the growth rate for $z < 2$. This is because the destruction rate (Equation \ref{eq:destruction_timescale}) includes the dust mass in its denominator; therefore the primary dust production channel heavily affects the destruction rate. 

The dust production rates from \citet{Popping17} (their Figure 10) and \citet{Vijayan19} (their Figure 8) show the same general trend as our model, in the sense that the net rate increases from high redshift to $z=2$ then decreases at later times. However, the rate from \citet{Popping17} is one order of magnitude higher than our model at the peak redshift. Their growth rate dominates the stellar production rate by more than 3 orders of magnitude at all redshift. The rate from \citet{Vijayan19} at the peak redshift is in the same order as ours. In their model, the production rate from SN II dominates at early times, switching over to grain growth at $z \sim 8$. Our model switches the dominance at around $z = 2$ from the stellar production to grain growth. These differences are due to different grain growth prescription and treatment for SN Ia. We use the dust growth prescription of \citet{Dwek1998} that does not account for the exchange of material between the diffuse ISM and molecular clouds. Both \citet{Popping17} and \citet{Vijayan19} use methods from \citet{Zhukovska14} that allows such treatment. In all models, the destruction rate closely follows the dominant production rate, showing the balance between production and destruction mechanisms of the interstellar dust.

\begin{figure*}
    \centering
    \includegraphics[width=0.8\textwidth]{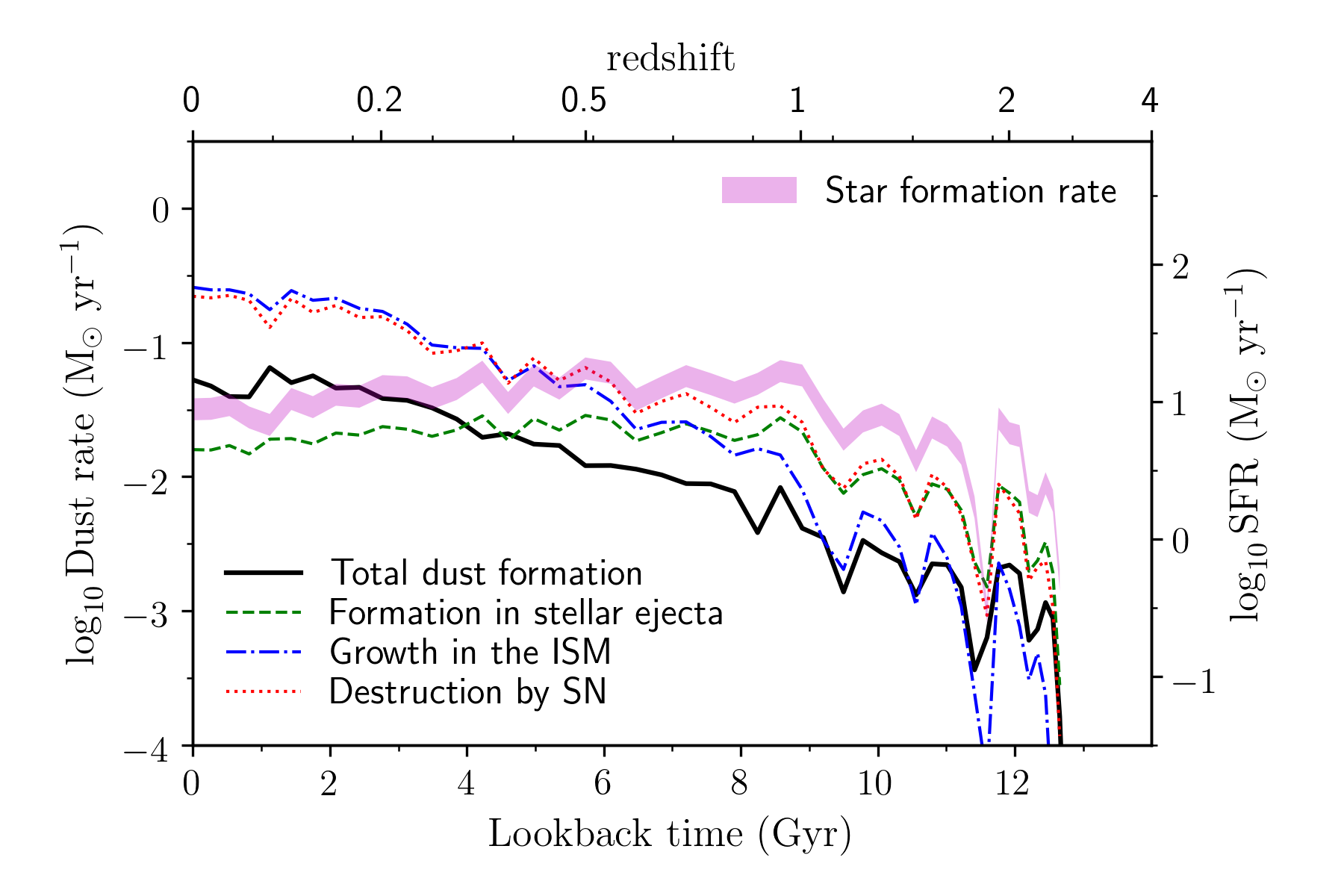}
    \caption{The dust and star formation history of a Miky Way-type model galaxy in \texttt{Dusty SAGE}. Thick solid black, green dashed, blue dash-dotted and red dotted lines represent the total dust rate, the dust formation rate in stellar ejecta, the dust growth rate in the ISM, and the dust destruction rate by SN, respectively. The magenta shaded regions shows the star formation rate density of our model, with the scale in the right axes.}
    \label{fig:individual}
\end{figure*}

Figure \ref{fig:individual} shows the dust and star formation history of a Milky Way-type model galaxy in \texttt{Dusty SAGE}. This galaxy has stellar mass $M_\mathrm{star} = 7.56 \times 10^{10} \Msun$, a bulge to stellar mass ratio = 0.354, $M_\mathrm{dust} = 1.99 \times 10^7 \Msun$ and an average star formation rate (SFR) = 8.84 \Msun $\mathrm{yr}^{-1}$. From the total rate density in Figure \ref{fig:DFRD} and the individual formation history in Figure \ref{fig:individual} we can see that the trend in the star formation history is followed consistently by the formation history of dust in the stellar ejecta at all redshifts. In Figure \ref{fig:individual}, both dust production and destruction channels follow the trend in star formation rate closely at $z > 6$ when the formation in stellar ejecta is the primary dust source. When the SFR slightly declines at $z < 6$, the formation rate in stellar ejecta also declines, but both the growth and destruction rates increase. We can see that the growth rate only starts picking up after there is a sufficient existing dust mass in the ISM. We also find that the destruction rate follows the rate of the dominant production channel, likely due to the dependence of destruction rate on the total dust mass in the ISM (Equation \ref{eq:destruction}). At early times until $z \sim 1$, the destruction rate follows the formation rate in stellar ejecta. After the dominant production channel shifts to the growth in the ISM, the destruction rate follows the growth rate.

\subsubsection{How dust production channels affect the DTM and DTG ratios}
\label{sssec:relation}
The trends in the DTG and DTM scaling relations follow each other. In the relations with gas phase metallicity, we find that there is a turning point at an oxygen abundance of $\sim8.5$, where the gradient of the relations change significantly. Below this point, the DTM ratio decreases with metallicity while the DTG ratio slightly increases. Above this point, both ratios steeply increase until they reach a point around $\sim 9.2$, after which the steepening stops and the slope becomes shallow again.

\begin{figure}
    \centering
    \includegraphics[width = 0.5\textwidth]{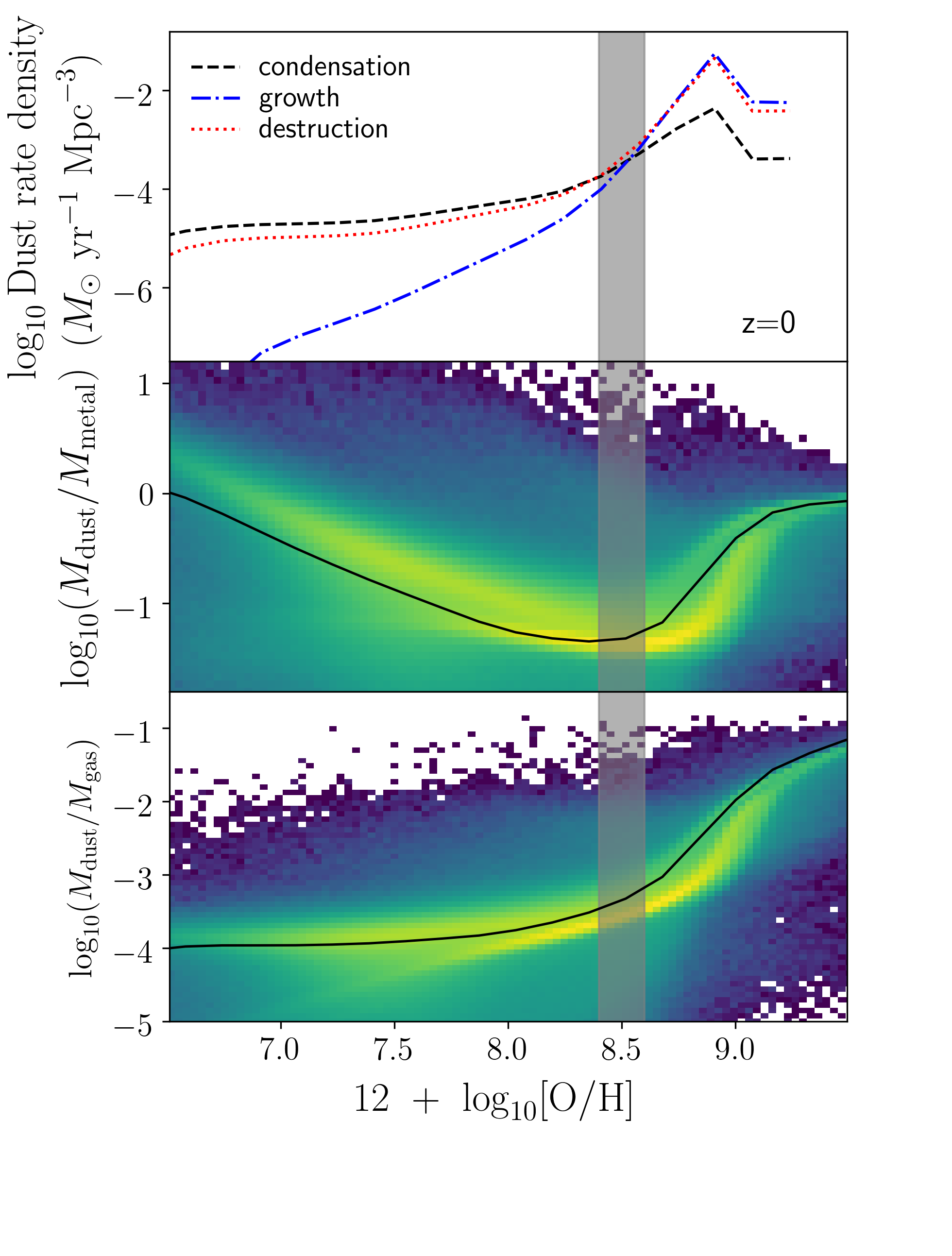}
    \caption{\textit{Top panel} shows our prediction of the dust production/destruction rate as a function of gas-phase metallicity at redshift $z=0$. Black dashed, blue dash-dotted and red dotted lines represent the rate density of dust formation in stellar ejecta, dust growth in the ISM, and dust destruction by SN, respectively. \textit{Middle panel} shows our prediction of the dust-to-metal (DTM) ratio as a function of gas phase metallicity at redshift $z=0$. \textit{Bottom panel} shows our prediction of the dust-to-gas (DTG) ratio as a function of gas phase metallicity at redshift $z=0$. In the middle and bottom panels, the heat map shows the 2D density distribution of galaxies in our model with brighter color representing higher density, while the black lines mark the median. The grey shaded region show the critical point in the production rate where growth starts to dominate.}
    \label{fig:7}
\end{figure}

We compare both relations with the dust production rate density in Figure \ref{fig:7} to investigate the cause for these features. The top panel presents the rate density of the three major processes that drive the dust content in our model: condensation in stellar ejecta, dust growth in molecular clouds, and destruction by SN shocks. The rate density is plotted against the same oxygen abundance range as the DTM and DTG ratios. This figure shows that the turning point of the gradient at oxygen abundance $\sim 8.5$ in the DTM and DTG ratio relations is also a turning point in the relation of dust production rate density with metallicity. Below this point, the condensation in the stellar ejecta dominates the overall dust production. As in this process a fixed fraction of the metals in stellar ejecta is converted to dust, the DTM and DTG ratios flatten. Above this point, the dust growth overtakes the other processes and increases steeply until $\sim 9.2$, causing the DTM and DTG ratios to rise sharply as well. Above $\sim 9.2$, the dust production and destruction rate decrease, corresponding to the shallow slope in the DTM and DTG ratios. 

\begin{figure}
    \centering
    \includegraphics[width = 0.5\textwidth]{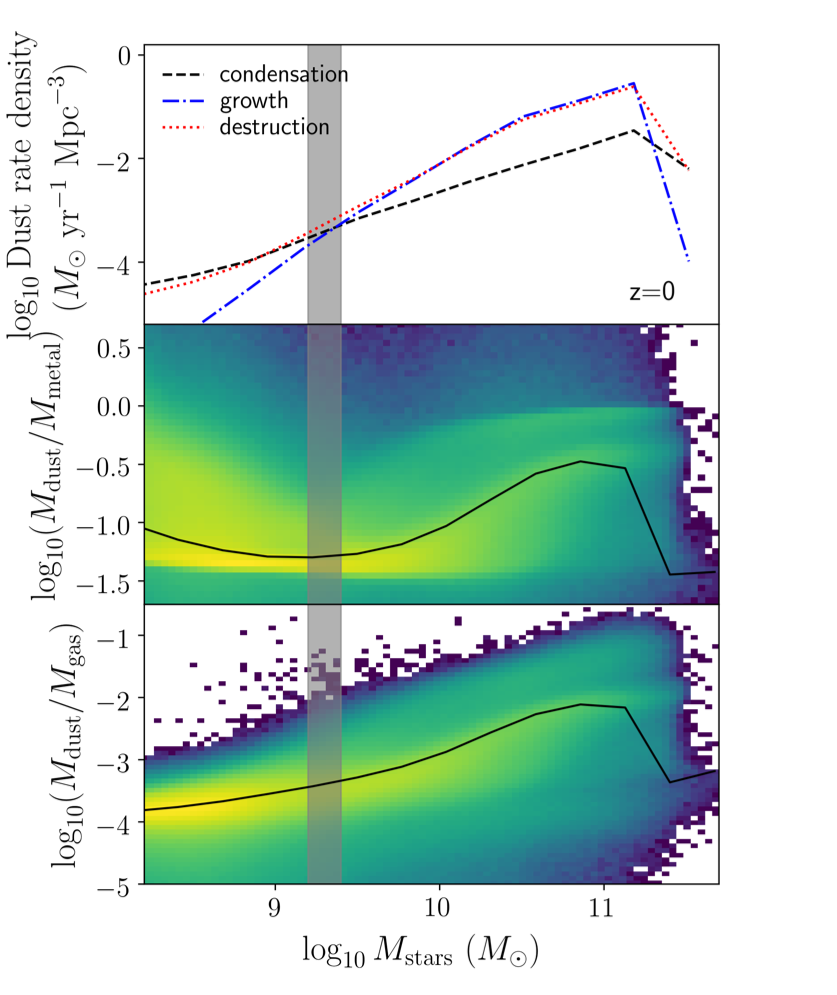}
    \caption{\textit{Top panel} shows our prediction of dust production/destruction rate as a function of stellar mass at redshift $z=0$. Black dashed, blue dash-dotted and red dotted lines represent the rate density of dust formation in stellar ejecta, dust growth in the ISM, and dust destruction by SN, respectively. \textit{Middle panel} shows our prediction of the dust-to-metal (DTM) ratio as a function of stellar mass at redshift $z=0$. \textit{Bottom panel} shows our prediction of the dust-to-gas (DTG) ratio as a function of stellar mass at redshift $z=0$. In the middle and bottom panels, the heat map shows the 2D density distribution of galaxies in our model with brighter color representing higher density, while the black lines mark the median. The grey shaded region show the critical point in the production rate where growth starts to dominate.}
    \label{fig:10}
\end{figure}

We show the same comparison against stellar mass in Figure \ref{fig:10}. The same connections with the dominant production process are again present. For example, for galaxies with stellar mass range $10^9 - 10^{11} \Msun$ at redshift $z=0$, where the dust growth is the main driver for dust production, we can see a sharp rise in both the DTG and DTM ratios. This relation shows that the increase in the DTM and DTG ratio almost always correlates with the dust growth in the molecular clouds. 

The correlation is further confirmed when we examine the evolution of the DTG and DTM scaling relations in Figures \ref{fig:5} to \ref{fig:9}. Here we see that above redshift $z=4$, there are no significant changes with stellar mass. But not so by redshift $z=3$, where we can see a gradual increase and steepening in all scaling relations towards the present day. These trends correspond to the behaviour seen in Figure \ref{fig:DFRD}, where we show the evolution of the total dust production density from redshift $z=7$ to $z=0$. This increase in the DTM and DTG scaling relations correlate with the redshift range where the growth in the ISM becomes the main production channel for dust. Using this correlation, we can infer the dust evolution in galaxies by investigating the trend in DTM and DTG ratios. 

In our condensation recipe, a fraction of molecular gas turns to stars, and when the stars die, they eject metals. A fraction of those metals then condenses into dust. Hence, it is natural that the DTG ratio is flat and independent of both metallicity and stellar mass in regions where the condensation dominates. In the dust growth recipe, the existing dust grains accrete metals in the dense ISM regions and grow their mass. Therefore, we expect a dependence between the DTG ratio and metallicity. The higher the metallicity, the more metals are available for the dust to grow, leading to higher DTG and DTM ratios. Given metallicity increases with stellar mass, we see the same trends in the DTG and DTM ratio versus stellar mass.

\subsection{Evolution of dust in the hot halo and ejected reservoir}
\label{sssec:hot dust}
\begin{figure*}
    \centering
    \includegraphics[width = 1.0\textwidth]{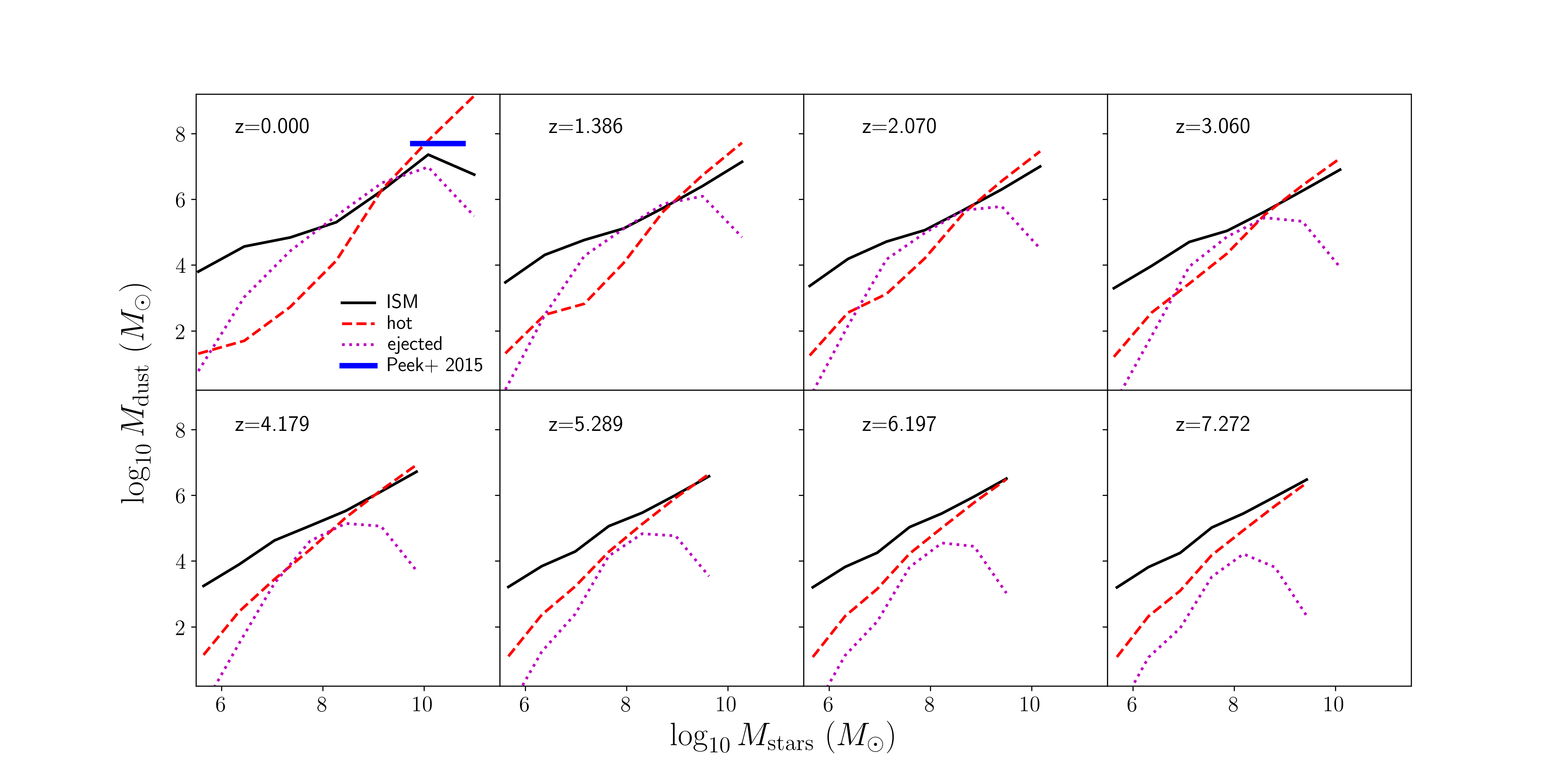}
    \caption{The mass of dust in the ISM (solid black line), hot halo (red dashed line) and ejected reservoirs (magenta dotted line) as a function of stellar mass of the host galaxy. At redshift $z=0$, we compare our prediction with the dust observations in the CGM from \citet{PMC15}.}
    \label{fig:12}
\end{figure*}

In our model, we track the gas transfer in three distinct mass reservoirs: ISM, hot halo and ejected. As gas is enriched with dust and metals, the movement of gas also implies their movement. We present these three dust mass components as a function of the stellar mass of the host galaxies in Figure \ref{fig:12}. The evolution of dust in the ISM has been described in the previous section. In this section, we will focus on the dust in the hot halo and ejected reservoir. 

Figure \ref{fig:12} shows that the dust mass in the hot halo increases with the stellar mass of the host galaxy. The gradient of the increase flattens above a stellar mass of $M_* = 10^{10} \Msun$. This flattening is also shown in \citet{Popping17}, where they attributed it to an increase of the destruction of hot halo dust by sputtering; the sputtering efficiency depends on halo virial temperature which correlates with host halo mass. 

We assume the dust mass in the hot halo corresponds to the dust mass in the circumgalactic medium (CGM), which is only partially true. With this, we can compare our predictions at redshift $z=0$ with the CGM dust mass observed by \citet{PMC15}. Our prediction match the observations only for galaxies with stellar mass $M_* \sim 10^{10} \Msun$. Above this mass, our model has a steeper slope compared to \citet{PMC15}. 

Figure \ref{fig:12} also shows an evolution in the relation of hot halo dust mass with stellar mass, which increases with time. The growth is more prominent for massive galaxies. Interestingly, at redshift $z<4$, the dust mass stored in the hot halo surpasses the dust mass in the ISM for the most massive galaxies. Since dust production only occurs in the ISM, the abundance of dust in the hot halo indicates a massive dust outflow from the ISM.

The dust mass accumulated in the ejected reservoir is lower than that in the ISM and the hot halo. This reservoir shows a distinct feature where it increases with stellar mass up to a critical mass, and then decreases rapidly. The overall relation between ejected dust mass and central stellar mass decreases gradually with lookback time. The critical mass also decreases from $M_* \sim 10^{8.1} \Msun$ to $M_* \sim 10^{10} \Msun$. Our finding is consistent with \citet{Popping17}. They argued that in galaxies above the critical mass, the heated dust and gas by SN feedback is trapped within the dark matter halo potential instead of flowing to the ejected reservoir, thus causing a decline.

\begin{figure}
    \centering
    \includegraphics[width = 0.5\textwidth]{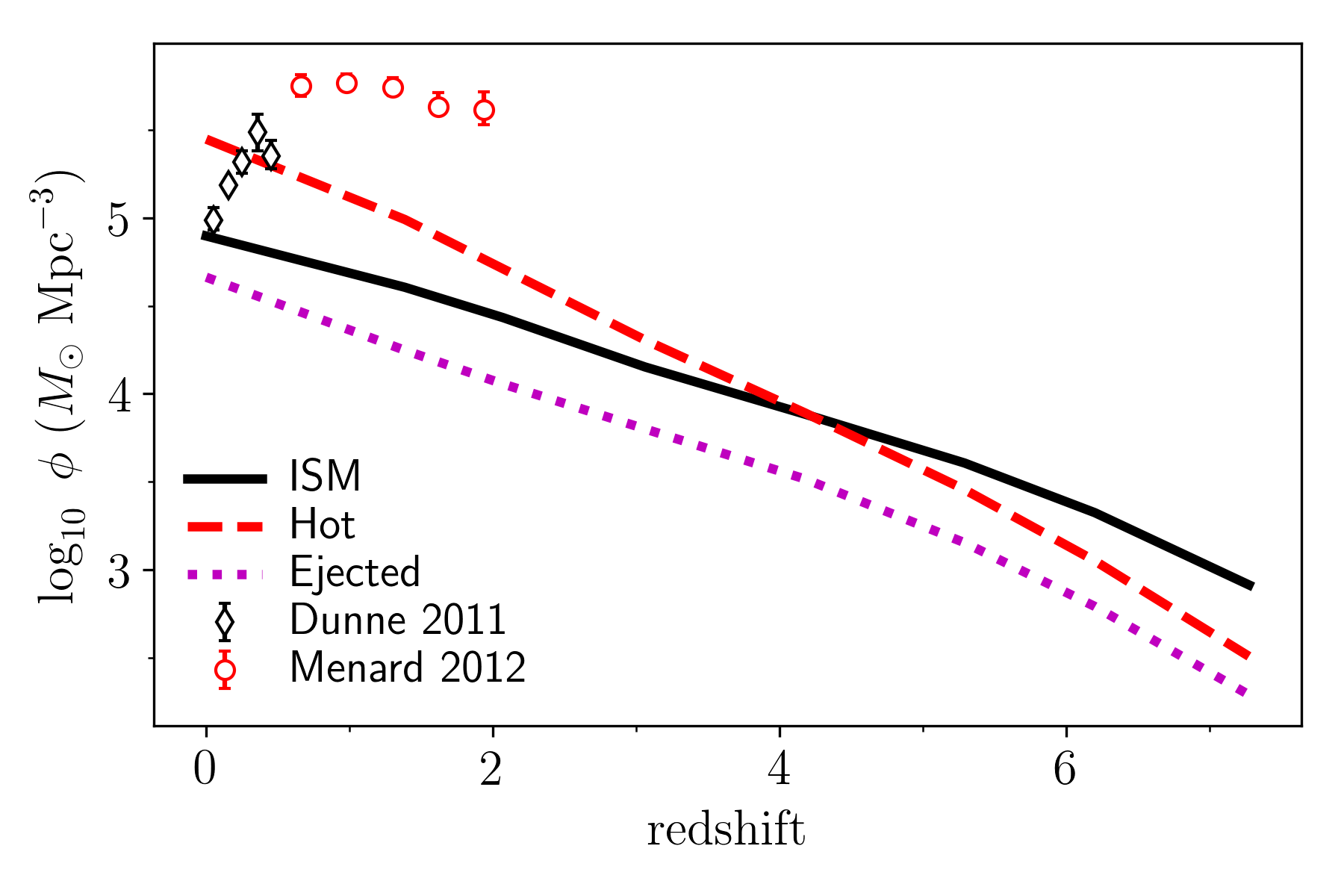}
    \caption{The cosmic density of dust in the ISM, hot halo and the ejected reservoir as a function of redshift. We compare our prediction of dust in the ISM with the observational constraints from \citet{Dunne11}, and in the hot halo with CGM observations of \citet{MF12}.}
    \label{fig:13}
\end{figure}

We present the cosmic density of dust mass in the ISM, hot halo and ejected reservoir in Figure \ref{fig:13}. Dust in all reservoirs increases with time. Initially, dust is produced in stellar ejecta and grows in the ISM. Therefore, most of the dust in galaxies is stored in the ISM at early times. During mergers and gas infall/outflows, this dust moves with the heated gas to the hot halo and ejected reservoir. Our model suggests that once the Universe has evolved below $z \sim 4$, there is more dust stored in the hot halo than in the ISM. The amount of dust in the ejected reservoir is the smallest across all cosmic time. 

Our predictions for ISM dust roughly matches the observations of \citet{Dunne11} at low redshift. Above $z = 0$, the observations are $0.2 - 0.5$ dex higher than our model. We compare our prediction for dust mass density in the hot halo with the CGM dust density observed by \citet{MF12}. Our prediction gives lower values but shows the same trend where the mass density increases with cosmic time.

\section{Conclusion}
\label{sec:conclusion}

Our model, \texttt{Dusty SAGE}, incorporates prescriptions for dust production and destruction into the \texttt{SAGE} semi-analytic galaxy model. We run this model on the Millennium simulation to make predictions for the dust content in galaxies. Our conclusions are the following:
\begin{itemize}
    \item Our model prediction for the dust mass function gives a good fit to the observations at redshift $z=0$ (Figure \ref{fig:11}). At higher redshift, our prediction is significantly lower than the observations. More work is needed to resolve these discrepancies.
    \item Our model for the dust mass - stellar mass relation is in good agreement with the observations at redshifts $z = 0$. At higher redshifts, our predictions indicate a lower dust mass compared to the observations. (Section \ref{sssec:dust-stellar mass}).
    \item Our predicted dust-to-gas (DTG) ratio increases with both gas phase metallicity and stellar mass, especially above a characteristic point where the slope becomes substantially steeper. Below this point, the DTG ratio is flat (Section \ref{sssec:dtg}).
    \item We find a characteristic point in the relation of the dust-to-metal (DTM) ratio with stellar mass and metallicity. We also find the DTM ratio increases sharply with both metallicity and stellar mass above this point (Section \ref{sssec:dtm}). 
    \item Our model predicts that in the early Universe, the major production channel of dust in the ISM is by its condensation in stellar ejecta. Between redshift $2 < z < 3$, grain growth in molecular clouds dominate to become the main driver of dust content until the present day. This growth is more significant for massive galaxies (see Section \ref{ssec:dominant shift}).
    \item The characteristic point in both the DTM and DTG relations correspond to where grain growth starts to dominate the dust production mechanism (Section \ref{sssec:relation}). This is a new insight. For the relations with gas phase metallicity, the characteristic point is around $12 + \log[O/H] = 8.5$, while for the stellar mass it is around $\log M_* = 9.2 \Msun$. Below this point, the slight decrease in the DTM relations and the flat DTG ratios are due to condensation in stellar ejecta. Above this point, the sharp increase in DTM and DTG ratios correspond to grain growth in dense molecular clouds. Hence, the scaling relations of the DTM and DTG ratios can be an indicator of the dominant dust production mechanism.
    \item We find a significant amount of dust in the hot halo and ejected reservoir. The hot halo is an important dust reservoir for massive galaxies. For redshifts below $z=3$, the integrated cosmic density of dust in the hot halo is larger than in the ISM. However, our predicted hot halo dust mass density is still 0.2 dex lower than the observed dust density in the CGM (Section \ref{sssec:hot dust}).
\end{itemize}

\texttt{Dusty SAGE} provides a set of predictions for future surveys of the dust content and their scaling relations in galaxies. Next generation telescopes such as ALMA and JWST are and will provide data in unprecedented detail, against which our predictions can be compared. In our model, the effect of dust on the physics of galaxy formation and evolution are still to be accounted for. We will improve this area in the future. \texttt{Dusty SAGE} is available at https://github.com/dptriani/dusty-sage.

\section*{Acknowledgements}
We would like to thank the referee for constructive feedback that helped to improve the paper. We would also like to thank Ned Taylor for helpful comments during the final stage of preparation. DPT would like to thank Anne Hutter for fruitful discussions regarding dust in galaxies, and Gergo Popping for providing data and for valuable suggestion. This research were supported by the Australian Research Council Centre of Excellence for All Sky Astrophysics in 3 Dimensions (ASTRO 3D), through project number CE170100013. The Semi-Analytic Galaxy Evolution (SAGE) model used in this work is a publicly available codebase that runs on the dark matter halo trees of a cosmological N-body simulation. It is available for download at https://github.com/darrencroton/sage.



\bibliographystyle{mnras}
\interlinepenalty=10000
\bibliography{bibliography} 







\bsp	
\label{lastpage}
\end{document}